\documentclass[final,1p,times]{elsarticle}
\usepackage{graphicx,amsmath,amsthm,amssymb,amsfonts,pstricks,pst-node,pst-plot,hyperref,algorithmic,pst-coil}
\usepackage{subfigure}
\usepackage{multirow}
\journal{Nuclear Physics B}
\usepackage[plain]{algorithm}

\theoremstyle{plain}
\newtheorem{theorem}{Theorem}[section]
\newtheorem{proposition}[theorem]{Proposition}
\newtheorem{lemma}[theorem]{Lemma}
\newtheorem{corollary}[theorem]{Corollary}
\newtheorem{algthm}{Algorithm}

\theoremstyle{definition}

\theoremstyle{remark}
\newtheorem{remark}{Remark}[section]

\begin{document}
\newcommand{\ontop}[2]{\genfrac{}{}{0pt}{}{#1}{#2}}
\newcommand{\bfsigma}{\boldsymbol{\sigma}}
\newcommand{\<}{\langle}
\renewcommand{\>}{\rangle}
\newcommand{\Tr}{{\rm Tr}}
\renewcommand{\active}{{\rm red}}
\newcommand{\inactive}{{\rm blue}}
\newcommand{\var}{{\rm var}}
\newcommand{\comm}{{\rightsquigarrow}}
\newcommand{\intercomm}{{\leftrightsquigarrow}}
\newcommand{\symdif}{\triangle}
\newcommand{\Dumbbell}{\mathcal{D}}
\newcommand{\Eulerian}{\mathcal{E}}
\newcommand{\Tadpole}{\mathcal{T}}
\newcommand{\pworm}{P_{G,n,x}}
\newcommand{\pwormprime}{P_{G,n,x}'}
\newcommand{\pwormbar}{\overline{P}_{G,n,x}}
\newcommand{\piworm}{\pi_{G,n,x}}
\newcommand{\piwormprime}{\pi_{G,n,x}'}
\newcommand{\piwormbar}{\overline{\pi}_{G,n,x}}
\newcommand{\pimodworm}{\pi_{G,n,x}'}
\newcommand{\pmodworm}{P_{G,n,x}'}
\newcommand{\pimodwormbar}{\overline{\pi'}_{G,n,x}}
\newcommand{\pmodwormbar}{\overline{P'}_{G,n,x}}
\newcommand{\XhOn}{X_{h}}
\newcommand{\XtPotts}{X_{P, t}}
\newcommand{\XtSubPotts}{X_{P, t2}}
\newcommand{\XhPotts}{X_{P, h}}
\newcommand{\lhs}{\text{LHS}}
\newcommand{\rhs}{\text{RHS}}
\newcommand{\pworminfty}{P_{G,n}}
\newcommand{\pwormloop}{\overline{\overline{P}}_{G,n,x}}
\newcommand{\pwormloopinfty}{\overline{\overline{P}}_{G,n}}
\newcommand{\pwormloopone}{\overline{\overline{P}}_{G,1,x}}
\newcommand{\pworminftyprime}{P_{G,n,\infty}'}
\newcommand{\pworminftyprelim}{P_{G,n}'}
\newcommand{\piworminfty}{{\pi}_{G,n}}
\newcommand{\piworminftyprelim}{{\pi}_{G,n}'}
\newcommand{\pworminftybar}{\overline{P}_{n}}
\newcommand{\pworminftybartwo}{\overline{P}_{2}}
\newcommand{\piworminftybar}{\overline{\pi}_{n}}
\newcommand{\pwormcolour}{P_{\rm colour}}
\newcommand{\pwormcolourbar}{\overline{P}_{\rm colour}}
\newcommand{\fp}{\mathcal{F}(G)}
\newcommand{\wormspace}{\mathcal{S}(G)}
\newcommand{\fpeven}{\mathcal{F}_0(G)}
\newcommand{\fpcolour}{\mathfrak{C}^{e,3}(H)}
\newcommand{\fpcoloureven}{\mathfrak{C}^{e,3}_0(H)}
\newcommand{\dualcolour}{\mathfrak{C}^{4}(H^*)}
\newcommand{\statespace}{\mathcal{S}}
\newcommand{\subspace}{\mathcal{X}}
\newcommand{\colour}{\boldsymbol{\sigma}}
\newcommand{\colourset}{\{1,2,\dots,m\}^V}
\newcommand{\pactive}{P_{\active}}
\newcommand{\pcolour}{P_{\rm colour}}
\newcommand{\pbond}{P_{\rm bond}}
\newcommand{\activeW}{\widehat{W}}
\newcommand{\edgeset}{A}
\newcommand{\jointmeasure}{\mu_{G,{\bf W}}}
\newcommand{\supp}{\text{supp}}

\setlength{\tabcolsep}{3pt}

\begin{frontmatter}

\title{Worm Monte Carlo study of the honeycomb-lattice loop model}

\author[ustc]{Qingquan Liu}
\ead{liuqq@mail.ustc.edu.cn}
\author[ustc]{Youjin Deng\corref{cor1}}
\ead{yjdeng@ustc.edu.cn}
\author[unimelb]{Timothy M. Garoni}
\ead{t.garoni@ms.unimelb.edu.au}
\cortext[cor1]{Corresponding author}
\address[ustc]{Hefei National Laboratory for Physical Sciences at Microscale,\\ Department of Modern Physics, University of Science and Technology of China,
  \\Hefei, 230027, China}
\address[unimelb]{ARC Centre of Excellence for Mathematics and Statistics of Complex Systems, \\
  Department of Mathematics and Statistics, The University of Melbourne, \\Victoria~3010, Australia}

\begin{abstract}
We present a Markov-chain Monte Carlo algorithm of {\em worm} type
that correctly simulates the $O(n)$ loop model on any (finite and connected) bipartite cubic graph, for any real $n>0$, and any edge weight, including the fully-packed limit of infinite edge weight.
Furthermore, we prove rigorously that the algorithm is ergodic and has the correct stationary distribution.
We emphasize that by using known exact mappings when $n=2$, this algorithm can be used to simulate a number of zero-temperature Potts antiferromagnets for
which the Wang-Swendsen-Koteck\'y cluster algorithm is non-ergodic, including the $3$-state model on the kagome lattice and the $4$-state model on the triangular lattice.
We then use this worm algorithm to perform a systematic study of the honeycomb-lattice loop model as a function of $n \le 2$, on the critical line and in the densely-packed and fully-packed phases.
By comparing our numerical results with Coulomb gas theory, we identify a set of exact expressions for
scaling exponents governing some fundamental geometric and dynamic observables.
In particular, we show that for all $n \le 2$, the scaling of a certain return time in the worm dynamics is governed by the magnetic dimension of the loop model,
thus providing a concrete dynamical interpretation of this exponent.
The case $n>2$ is also considered, and we confirm the existence of a phase transition in the 3-state Potts universality class that was recently observed via numerical transfer matrix calculations.
\end{abstract}

\date{November 9, 2010}
\end{frontmatter}

\section{Introduction}
\label{intro}
Among the myriad of models studied in the theory of critical phenomena, two fundamental examples that continue to play a central role are the $q$-state Potts model~\cite{Potts52,Wu82,Wu84},
and the $O(n)$ spin model~\cite{Stanley68,PelissettoVicari02}.
In the original spin representation, the parameter $q$ or $n$ must be a positive integer. However, the Fortuin-Kasteleyn representation~\cite{FortuinKasteleyn72}
of the ferromagnetic Potts model and the loop representation~\cite{DomanyMukamelNienhuisSchwimmer81} of the $O(n)$ spin model show how these models can be extended
to arbitrary real $q, n \ge 0$, by re-expressing them as models of random geometric objects: clusters or loops, respectively.
In fact, the extension of the Potts model to non-integer $q$ can also be formulated directly in the spin language, by re-expressing the Potts spin clusters in terms of domain walls~\cite{DubailJacobsenSaleur10}.
These geometric models play a major role in recent developments of conformal field theory~\cite{DiFrancescoMathieuSenechal97}
via their connection with Schramm-Loewner evolution (SLE)~\cite{KagerNienhuis04,Cardy05}.

Monte Carlo methods are an indispensable tool in statistical mechanics~\cite{LandauBinder05,SokalLectures}.
The Sweeny algorithm~\cite{Sweeny83} and the Swendsen-Wang-Chayes-Machta cluster-algorithm~\cite{SwendsenWang87,ChayesMachta98} provide remarkably
efficient~\cite{DengGaroniSokal07b,DengGaroniMachtaOssolaPolinSokal07} tools for studying the ferromagnetic Potts (random-cluster~\cite{Grimmett06}) model,
and are valid for any real $q>0$, or $q>1$, respectively.
For loop models, by contrast, efficient simulation at noninteger $n$ has posed a significant challenge.
Instead, numerical transfer-matrix techniques have typically been used~\cite{BloeteNienhuis89,GuoBloeteWu00}.
Monte Carlo simulations at $n \neq 1$ have been reported in~\cite{KarowskiThunHelfrichRys83,DingDengGuoQianBloete07},
however the algorithms used were essentially single-spin-flip Metropolis algorithms for Ising spins on the dual lattice.
As such, their efficiency is limited, and they are manifestly non-ergodic\footnote{Following the typical usage in the physics literature, we take {\em ergodic} as synonymous with {\em irreducible}.
Recall that a Markov chain is irreducible if for each pair of states $i$ and $j$ there is a positive probability that starting in $i$ we eventually visit $j$, and vice versa.}
at infinite edge weight.
In~\cite{DengGaroniGuoBloeteSokal07}, a cluster algorithm was presented that is valid for all $n\ge1$, which is dramatically more efficient than the single-spin-flip algorithms on the critical line.
However, its efficiency deteriorates rapidly as the edge weight increases, and it too becomes non-ergodic at infinite edge weight.

In~\cite{ZhangGaroniDeng09}, a Monte Carlo algorithm of {\em worm} type for simulating the honeycomb-lattice fully-packed loop model with $n=1$ was presented, and its validity was rigorously proved.
In this article, we present a worm algorithm that correctly simulates the $O(n)$ loop model on any bipartite
cubic graph,\footnote{All graphs considered in this article are implicitly assumed to be finite and connected.} for any real $n>0$, and any edge weight, including the fully-packed limit of infinite edge weight.
Furthermore, we prove rigorously that the algorithm is ergodic and has the correct stationary distribution.
We then use this algorithm to perform a systematic study of the honeycomb-lattice loop model as a function of $n$.
By comparing our numerical results for $n\le2$ with Coulomb-gas theory~\cite{Nienhuis84}, we identify the exact scaling exponents of some fundamental geometric observables, as well as certain
observables related to dual Ising spins.
Furthermore, we find that for all $n\le2$, the scaling dimension of a certain very natural return time in the worm dynamics coincides precisely with the magnetic dimension of the loop model,
which provides a concrete dynamical interpretation of this exponent which is meaningful for both integer and non-integer $n$.
See section~\ref{numerical section}.
We also study the case $n>2$, and confirm the existence of a critical transition in the 3-state Potts universality class, which was recently observed using transfer matrices~\cite{GuoBloeteWu00}.
While the honeycomb-lattice model is perhaps the archetypal loop model, and is certainly the most well-studied case,
there are other examples of bipartite cubic graphs which are of interest, including the $(4\cdot8^2)$ Archimedean lattice (dual of the Union Jack lattice),
and the Hydrogen-peroxide lattice (which is three-dimensional).
Systematic studies of the loop models on both of these lattices can be performed using the algorithms described in this article; the results will be presented elsewhere.

In addition to the study of loop models, the worm algorithms that we present here can also be applied to the study of a number of antiferromagnetic Potts models.
It is well known that the honeycomb-lattice fully-packed loop model with $n=1$ is equivalent to the zero-temperature triangular-lattice antiferromagnetic Ising model.
The latter model (which is critical) provides a canonical example of geometric frustration, but is notoriously difficult to simulate.
In fact, even the most sophisticated tailor-made cluster algorithms~\cite{ZhangYang94,CoddingtonHan94} are thought to be non-ergodic.
However, the worm algorithm constructed in~\cite{ZhangGaroniDeng09} immediately provides a provably ergodic Monte Carlo method for this problem.
Furthermore, it is known that the honeycomb-lattice fully-packed loop model with $n=2$ is equivalent to both the zero-temperature kagome-lattice 3-state Potts antiferromagnet and
the zero-temperature triangular-lattice 4-state Potts antiferromagnet~\cite{Baxter70}. Both of these models are believed to be critical.
While the Wang-Swendsen-Koteck\'y~\cite{WangSwendsenKotecky90} (WSK) cluster algorithm is undoubtedly the current state-of-the-art for simulating antiferromagnetic Potts models,
it has recently been proved~\cite{MoharSalas09,MoharSalas10} to be non-ergodic for both of these cases.
By contrast, the worm algorithms described in Section~\ref{Worm algorithms} have been proved to be ergodic and
can be applied in a straightforward way to the study of both of these Potts antiferromagnets.
The details of this application will be reported elsewhere (but see also Section~\ref{discussion} for further discussion).

\begin{figure}
  \caption{\label{loop figures} Typical loop configuration \subref{loop figure} and fully-packed loop configuration \subref{FPL figure} on the honeycomb lattice with periodic boundary conditions.
    Thick lines denote occupied edges, thin lines denote vacant edges.}
  \begin{center}
    \psset{unit=0.4cm}
    \subfigure[Loop configuration]{\label{loop figure}
      \includegraphics{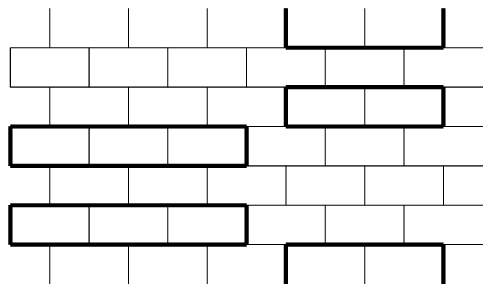}
    }
    \subfigure[Fully-packed loop configuration]{\label{FPL figure}
      \includegraphics{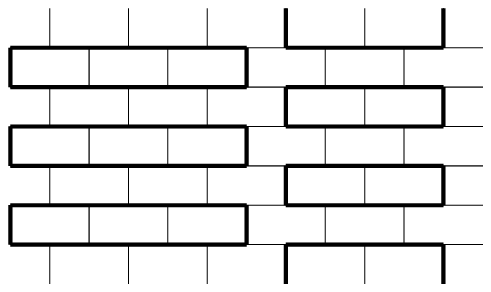}
    }
  \end{center}
\end{figure}

Worm algorithms were first applied to classical lattice models in~\cite{ProkofevSvistunov01}, and it was demonstrated empirically in~\cite{DengGaroniSokal07c} that the worm algorithm is
an extraordinarily efficient method for simulating the three-dimensional Ising model. See~\cite{JankeNeuhausSchakel10,Wolff10a} for some recent applications to $O(n)$ models.
Worm algorithms provide a natural way to simulate {\em cycle-space} models.
Given a finite graph $G=(V,E)$, the cycle space, $\mathcal{C}(G)$, is the set of all $A\subseteq E$ such that every vertex in the spanning subgraph $(V,A)$ is even.
We call $A\subseteq E$ and $(V,A)$ {\em Eulerian} whenever $A\in\mathcal{C}(G)$.
Fig.~\ref{loop figures}\subref{loop figure} shows a typical configuration on the honeycomb lattice.
The essence of the worm idea is to enlarge the state space $\mathcal{C}(G)$ to include a pair of {\em defects} (i.e., vertices of odd degree), and then to move these defects via random walk.
When the two defects collide, the configuration becomes Eulerian once more.
A very natural class of cycle-space models is defined for $n,x>0$ by the probability measure
\begin{equation}
\phi_{G,n,x}(A) \propto \, n^{c(A)}\,x^{|A|},\qquad A\in\mathcal{C}(G),
\label{loop measure}
\end{equation}
where $c(A)$ is the cyclomatic number of the spanning subgraph $(V,A)$.
The cyclomatic number of a graph is simply the minimum number of edges to remove from it in order to make it cycle-free.
On graphs $G$ of maximum degree $\Delta(G)\le3$ therefore, all $A\in\mathcal{C}(G)$ consist of a collection of disjoint cycles, or {\em loops}, and $c(A)$ is then simply the number of such loops.
Consequently, the model \eqref{loop measure} is typically referred to as the {\em loop model}, and bipartite cubic graphs (such as the honeycomb lattice) provide a natural setting for its
study.

It is well known~\cite{DomanyMukamelNienhuisSchwimmer81} that on any graph $G=(V,E)$ of maximum degree $\Delta(G)\le\,3$,
the model \eqref{loop measure} arises for positive integer $n$ as a loop representation of
an $n$-component spin model,
\begin{equation}
  Z = \Tr\,\prod_{ij\in E}\,(1+n\,x\,\bfsigma_i\cdot\bfsigma_j),
  \label{spin model}
\end{equation}
where $\bfsigma=(\sigma^1,\dots,\sigma^n)\in\mathbb{R}^n$ and $\Tr$ denotes normalized integration with respect to any {\em a priori} measure $\<\cdot\>_0$ on $\mathbb{R}^n$
satisfying $\<\sigma^{\alpha}\sigma^{\beta}\>_0=\delta_{\alpha,\beta}/n$ and $\<\sigma^{\alpha}\>_0=\<\sigma^{\alpha}\sigma^{\beta}\sigma^{\gamma}\>_0=0$. In particular, uniform measure on the unit sphere
is allowed, as are various {\em face-cubic} and {\em corner-cubic} measures~\cite{DomanyMukamelNienhuisSchwimmer81}. For $n\neq1$, the Boltzmann weight \eqref{spin model} with spins on a sphere defines a
nonstandard $O(n)$ spin model, which has positive weights only for $|x|<1/n$, but it is, nevertheless, expected to belong to the usual $O(n)$ universality class.

In the limit $x\to+\infty$, the support of $\phi_{G,n,x}$ reduces to the set of all $A\in\mathcal{C}(G)$ with maximal $|A|$.
On bipartite cubic graphs $\max_{A\in\mathcal{C}(G)}|A|=|V|$, so the set of all such {\em fully-packed} configurations is simply
\begin{equation}
\fp:=\{A\in\mathcal{C}(G)\,:\,d_v(A)=2 \text{ for all } v\in V\}.
\label{bipartite cubic fp space}
\end{equation}
Fig.~\ref{loop figures}\subref{FPL figure} shows a typical fully-packed configuration on the honeycomb lattice.
We note that the elements of $\fp$ are referred to as {\em $2$-factors}\footnote{More precisely, if $A\in\fp$ then the spanning subgraph $(V,A)$ is a 2-factor.} by graph theorists~\cite{Diestel05}.
We also remark that $A\in\fp$ iff $E\setminus A$ is a dimer covering (perfect matching) of $G$.
Finally, we note that the limiting measure is simply
\begin{equation*}
\phi_{G,n}(A) := \phi_{G,n,\infty}(A)
\propto
n^{c(A)},\qquad A\in \fp.
\end{equation*}

A great deal is known about the loop model \eqref{loop measure} on the honeycomb lattice when $n\le 2$.
For given $n$, the model is believed to have three distinct phases: a disordered phase (small $x$), a densely-packed (DP) phase (large finite $x$), and a fully-packed (FP) phase (infinite $x$).
Furthermore, the model is exactly solvable~\cite{Baxter86,Baxter87,BatchelorBloete88,BatchelorBloete89,Suzuki88} on the curves
\begin{equation}
x = \frac{1}{\sqrt{2\pm\sqrt{2-n}}}.
\label{integrable curve}
\end{equation}
The plus sign in \eqref{integrable curve} corresponds to the critical curve, $x_c(n)$, separating the disordered and densely-packed phases~\cite{Nienhuis82}.
The minus sign in \eqref{integrable curve} corresponds to a curve of stable fixed points in the densely-packed phase.
For all finite $x>x_c$, the loop model is in the densely-packed phase, which is critical~\cite{BloeteNienhuis89}.
The fully-packed model is also critical, however it is known to be in a distinct universality class to the densely-packed phase~\cite{BloeteNienhuis94,BatchelorSuzukiYung94,KondevDeGierNienhuis96}.
For convenience, we shall refer to the model \eqref{loop measure} with $x_c<x<\infty$ as the densely-packed loop (DPL) model, and to the $x=\infty$ model as the fully-packed loop (FPL) model.

The loop model with loop fugacity $n$ can be related to a Coulomb gas~\cite{Nienhuis84,KondevDeGierNienhuis96} with coupling $g$ by
\begin{equation}
  n = - 2 \cos(\pi g/4),
  \label{n-g relation}
\end{equation}
with
\begin{equation}
  g \in
  \begin{cases}
    [2,4], & x_c < x \le +\infty,\\
    [4,6], & x = x_c.
  \end{cases}
  \label{x-g relation}
\end{equation}
Recall~\cite{Nienhuis91} that the critical (densely-packed) loop model with $0\le n \le 2$ corresponds to a tricritical (critical) Potts model with $q = n^2$.
The normalization for $g$ given in \eqref{n-g relation} and \eqref{x-g relation}, which is a factor of 4 times larger than the $g$ presented in~\cite{Nienhuis84},
is in fact the standard normalization for the Coulomb gas corresponding to the $q=n^2$ Potts model, rather than the $O(n)$ loop model.
This choice facilitates easy translation between loop and Potts exponents, which will prove convenient in Section~\ref{numerical section}.

Coulomb gas theory~\cite{Nienhuis84} predicts a whole spectrum of exact scaling dimensions characterizing the loop model.
However, we emphasize that identifying {\em which} loop-model observables these exponents actually govern is not always obvious.
In~\cite{DengGaroniGuoBloeteSokal07}, Monte Carlo simulations were combined with Coulomb gas predictions to identify the exact scaling exponents for a number of natural geometric observables,
as well as observables related to dual Ising spins.
The results presented in~\cite{DengGaroniGuoBloeteSokal07}, however, were restricted to the critical branch, $x=x_c$, and to $n\ge1$.
In this work we shall use worm algorithms to extend these observations to all $n>0$, and to the DP and FP phases. See Section~\ref{numerical section}.

The outline of this article is as follows. The necessary theoretical results concerning the algorithms appear in Section~\ref{Worm algorithms}.
In Section~\ref{numerical section} we then present our numerical results for the $n\le2$ honeycomb-lattice loop model
in the critical, densely-packed, and fully-packed phases, and also for the $n>2$ model. Section~\ref{discussion} then concludes with a discussion.

\section{Worm dynamics for loop models}
\label{Worm algorithms}
We begin our discussion of worm dynamics by constructing a worm algorithm to simulate \eqref{loop measure} on an arbitrary graph, for any $0<n,\,x<\infty$.
This essentially generalizes the presentation in~\cite{DengGaroniSokal07c,ZhangGaroniDeng09} to include a loop fugacity in the stationary distribution.
We then demonstrate that it is possible to make this algorithm {\em rejection-free}, in a certain sense.
We then turn our attention to the specific case of cubic graphs, and consider the limit $x\to\infty$.
In particular, we rigorously prove that the rejection-free worm algorithm on any bipartite cubic graph remains ergodic at $x=+\infty$.

\subsection{Simple worm dynamics}
\label{standard worm}
Fix a finite graph $G=(V,E)$, and for any $A\subseteq E$ let $\partial A\subseteq V$ denote the set of all vertices which have odd degree in the spanning subgraph $(V,A)$.
Loosely, $\partial A$ is just the set of sites that {\em touch} an odd number of the bonds in the bond configuration $A$.
If $u,v\in V$ are distinct we write
\begin{equation*}
\mathcal{S}_{u,v}(G):=\{A\subseteq E\,:\, \partial A=\{u,v\}\},
\end{equation*}
and
\begin{equation*}
\mathcal{S}_{v,v}(G):=\{A\subseteq E\,:\,\partial A=\emptyset\}.
\end{equation*}
We emphasize that $\mathcal{S}_{v,v}(G)=\mathcal{C}(G)$ for every $v\in V$.
We take the state space of the worm algorithm to be
\begin{equation*}
\wormspace:=\{(A,u,v) \,:\, u,v\in V \text{ and } A\in \mathcal{S}_{u,v}(G)\},
\end{equation*}
i.e., all ordered triples $(A,u,v)$ with $A \subseteq E$ and $u,v \in V$, such that $A~\in~\mathcal{S}_{u,v}(G)$.
Note that if $(A,u,v)\in\wormspace$ then $A\in\mathcal{C}(G)$ iff $u=v$.
Thus, the bond configurations allowed in the state space of the worm algorithm constitute a superset of the Eulerian configurations.
Finally, we assign probabilities to the configurations in $\wormspace$ according to
\begin{equation}
\piworm(A,u,v)\propto  d_u\,d_v\, n^{c(A)}\,x^{|A|},\qquad (A,u,v)\in \wormspace,
\label{worm measure}
\end{equation}
where $d_v$ denotes the degree in $G$ of $v\in V$. In the following, when we wish to refer to the degree of $v\in V$ in the
spanning subgraph $(V,A)$ we will write $d_v(A)$. Loosely, $d_v(A)$ is simply the number of bonds that touch $v$ in the bond configuration $A$.
In this notation we have $d_v=d_v(E)$.

The first step in constructing the standard worm algorithm is to consider
the worm {\em proposal matrix}, $P^{(0)}$, which is defined for all $uu'\in E$ and $v\in V$ by
\begin{equation}
P^{(0)}[(A,u,v)\to(A\triangle uu',u',v)] = P^{(0)}[(A,v,u)\to(A\triangle uu',v,u')] = \frac{1}{2d_u},
\label{worm proposal}
\end{equation}
all other entries being zero. Here $\triangle$ denotes symmetric difference, i.e. delete the bond $uu'$ from $A$ if it is present, or insert it if it is absent.
It is easy to see that $P^{(0)}$ is an ergodic transition matrix on $\wormspace$.
According to~(\ref{worm proposal}), the moves proposed by the worm algorithm are as follows:
Pick uniformly at random one of the two defects (say, $u$) and one of
the edges emanating from $u$ (say, $uu'$), then move from the current configuration $(A,u,v)$ to the new configuration $(A~\triangle~uu',u',v)$.

Now we simply apply the usual Metropolis-Hastings prescription (see e.g.~\cite{SokalLectures}) to
assign acceptance probabilities to the moves proposed by $P^{(0)}$, so that the resulting transition matrix, $\pworm$, is in detailed balance with (\ref{worm measure}).
Explicitly, for all $uu'\in E$ and $v\in V$ we have
\begin{multline}
   \pworm[(A,u,v)\to(A\triangle uu',u',v)] = \pworm[(A,v,u)\to(A\triangle uu',v,u')] \\
= \frac{1}{2d_u}
\begin{cases}
  \min(1,x\,n)   & uu'\not\in A \text{ {\rm and} } u     \leftrightarrow u' \text{ {\rm in} } (V,A)             \\
  \min(1,x)      & uu'\not\in A \text{ {\rm and} } u \not\leftrightarrow u' \text{ {\rm in} } (V,A)             \\
  \min(1,1/n\,x) & uu'    \in A \text{ {\rm and} } u     \leftrightarrow u' \text{ {\rm in} } (V,A\setminus uu')\\
  \min(1,1/x)    & uu'    \in A \text{ {\rm and} } u \not\leftrightarrow u' \text{ {\rm in} } (V,A\setminus uu')
\end{cases}
\label{worm P}
\end{multline}
The notation $u\leftrightarrow u'$ in \eqref{worm P} means that vertices $u$ and $u'$ are connected in the stated spanning subgraph of $G$.
The transitions (\ref{worm P}) define $\pworm$ uniquely since all other transitions occur with zero probability, except the identity transitions $(A,u,v)\to(A,u,v)$,
whose transition probabilities are fixed by normalization.

Now let us return to our original goal, which was to sample from $\mathcal{C}(G)$.
As elaborated below in Lemma~\ref{sub-chain}, achieving this is as simple as running the worm chain and choosing to only measure observables when the two defects meet, $u=v$.
This defines an ergodic Markov sub-chain on
\begin{equation}
\{(A,u,v)\in\wormspace:u=v\} \cong \mathcal{C}(G)\times V
\label{eulerian subspace}
\end{equation}
with stationary distribution $\piwormbar(A,v)\propto n^{c(A)}\,x^{|A|}$, and therefore for any loop-model observable $X:\mathcal{C}(G)\to\mathbb{R}$ we
have $\langle X \rangle_{\piwormbar} = \langle X \rangle_{\phi_{G,n,x}}$.

The resulting Monte Carlo algorithm is summarized in Algorithm~\ref{simple worm algorithm}.
Note that the acceptance probabilities (which are simply $2\,d_u\,\pworm[(A,u,v)\to(A\symdif uu',u',v)]$) will in general depend on the topology of the loops in a non-trivial way;
we shall return to this point in Section~\ref{connectivity and colouring section}.
The abbreviation UAR simply means {\em uniformly at random}.
\begin{algorithm}
  \begin{algthm}[Simple worm algorithm] $\,$
    \label{simple worm algorithm}
    \begin{algorithmic}
      \LOOP
      \STATE Current state is $(A,u,v)$
      \STATE UAR, pick one of the two defects (say $u$)
      \STATE UAR, pick a neighbor $u'$ of $u$
      \STATE Make the transition $(A,u,v)\to(A\symdif uu',u',v)$ with acceptance probability inferred from \eqref{worm P}
      \IF{$u'=v$}
      \STATE Measure observables
      \ENDIF
      \ENDLOOP
    \end{algorithmic}
  \end{algthm}
\end{algorithm}

\begin{remark}
  The naive $n\to0$ limit of~\eqref{loop measure} with $x<\infty$ held fixed reduces to a trivial measure concentrated on the single state $A=\emptyset$,
  and the naive $n\to0$ limit of Algorithm~\ref{simple worm algorithm} leads to a trivial dynamics which correctly simulates this trivial model.
  Non-trivial $n\to0$ limits can be taken however; for example, conditioning on positive cyclomatic number before taking the $n\to0$ limit of \eqref{loop measure} yields a model of self-avoiding polygons.
  Worm dynamics can be developed to simulate both self-avoiding walks and self-avoiding polygons, however a discussion of these issues would lead us too far afield here.
  A discussion of such algorithms will be reported elsewhere.
\end{remark}

\subsection{Markov sub-chains}
Let us pause momentarily, and consider, quite generally, that we have an ergodic Markov chain on a finite state space $\statespace$ which is in detailed balance with a distribution $\pi$,
and suppose that we only observe the process when it is in a state in $\subspace\subset\statespace$.
This new process is a Markov chain on $\subspace$ with transition matrix
$$
(\overline{P})_{ss'}
:= (P)_{ss'} + \sum_{n=0}^{\infty} \,\sum_{s_0,s_1,\dots,s_n\in\overline{\subspace}}\,(P)_{ss_0}\,\prod_{l=1}^n\,(P)_{s_{l-1}s_l}\, (P)_{s_n s'}.
$$
\begin{lemma}
\label{sub-chain}
$\overline{P}$ is ergodic and in detailed balance with the restriction of $\pi$ to $\subspace$
$$
\overline{\pi}_s = \frac{\pi_s}{\sum_{s'\in \subspace}\pi_{s'}}, \qquad s\in\subspace.
$$
\begin{proof}
The ergodicity of $\overline{P}$ on $\subspace$ follows immediately from the ergodicity of $P$ on $\statespace$,
and one can easily verify directly that $\overline{P}$ is in detailed balance with $\overline{\pi}$ by using the fact that $P$ is in detailed balance with $\pi$.
\end{proof}
\end{lemma}

We now wish to make the following observation.
Since we only observe the $\statespace$-chain when it visits $\subspace\subset\statespace$, we have quite a bit of freedom to modify the transition probabilities
in $\overline{\subspace}=\statespace\setminus\subspace$ without affecting the stationary distribution of the $\subspace$-chain. In particular, we can forbid identity transitions in $\overline{\subspace}$.
\begin{corollary}
\label{rejection free corollary}
If
\begin{equation*}
(P')_{ss'}
:=
\begin{cases}
(P)_{ss'} & s\in \subspace\\
\displaystyle
\frac{(P)_{ss'}}{1-(P)_{ss}} & s\in \overline{\subspace}, s'\neq s\\
0 & s\in \overline{\subspace}, s'= s\\
\end{cases}
\end{equation*}
then $\overline{P'}$ is in detailed balance with $\overline{\pi}$.
\begin{proof}
It is elementary to verify directly that $P'$ is in detailed balance with
\begin{equation*}
\pi_s'=
\begin{cases}
\pi_s & s\in \subspace,\\
(1-(P)_{ss})\,\pi_s & s\in \overline{\subspace}.\\
\end{cases}
\end{equation*}
Lemma \ref{sub-chain} then immediately implies that $\overline{P'}$ is in detailed balance with $\overline{\pi}$.
\end{proof}
\end{corollary}
Therefore, both $\overline{P}$ and $\overline{P'}$ are in detailed balance with the same distribution $\overline{\pi}$.
Since $P'$ forbids identity transitions $s\to s$ when $s\in\overline{\subspace}$ (which correspond to rejections in the context of Metropolis algorithms)
$P'$ is clearly more efficient than $P$ at sampling from $\subspace$.
We shall refer to $P'$ as a rejection-free chain, although it should be emphasized that rejections are still allowed inside $\subspace$.

A concrete example of the advantage of using the rejection-free chain is provided by considering worm algorithms for fully-packed loop models.
Indeed, as we shall see, the $x\to\infty$ limit of $\pworm$ is absorbing, while the corresponding rejection-free algorithm remains ergodic (at least on bipartite cubic graphs).

\subsection{Rejection-free worm dynamics}
\label{rejection-free section}
Thus far we have glossed over an important issue, namely the ergodicity of $\pworm$.
It is not hard to see that $\pworm$ is ergodic whenever $x<\infty$.
However, $\pworm$ is manifestly non-ergodic when $x=+\infty$, in fact it is absorbing. Indeed, as $x\to\infty$ the probabilities for transitions that remove an edge vanish.
Consequently,
\begin{equation*}
  P_{n,+\infty}[(A,u,v)\to(A,u,v)]
  =
  1 - \frac{d_u - d_u(A)}{2\,d_u} - \frac{d_v - d_v(A)}{2\,d_v},
\end{equation*}
and all states $(A,u,v)\in\wormspace$ for which both $d_u(A)=d_u$ and $d_v(A)=d_v$ become absorbing as $x\to\infty$.

Suppose now that $G$ is $k$-regular. Then $(A,u,v)$ will be absorbing when $x=+\infty$ iff $d_u(A)=d_v(A)=k$.
By definition, if $(A,u,v)\in\wormspace$ then when $u=v$ the vertex degree $d_u(A)=d_v(A)$ is even, whereas when $u\neq v$ both $d_u(A)$ and $d_v(A)$ are odd.
Thus, if $k$ is odd then $(A,u,v)$ can be absorbing only if $u\neq v$, and so all states with $A\in\mathcal{C}(G)$ remain non-absorbing.
In particular, on a cubic graph we can now see that as $x\to \infty$ all states $(A,v,v)\in\mathcal{C}(G)\times V$ remain non-absorbing while all states
$(A,u,v)$ with $u\neq v$ and $d_u(A)=d_v(A)=3$ become absorbing; once both defects have degree $3$, the $\pworm$ chain remains in the given state for eternity.

We cannot, therefore, use $\pworm$ to simulate the fully-packed loop model. As we shall see, however, we can use its rejection-free counterpart.
Following Corollary~\ref{rejection free corollary}, we define a new transition matrix $\pwormprime$ by explicitly conditioning on making a non-trivial transition whenever $u\neq v$.
Specifically, we define
\begin{equation}
\begin{split}
\pwormprime[(A,u,v)\to (A\symdif uu',u',v)] &= \pwormprime[(A,v,u)\to (A\symdif uu',v,u')]\\
&=
\begin{cases}
\pworm[(A,u,v)\to (A\symdif uu',u',v)] & u=v,\\
\\
\displaystyle
\frac{\pworm[(A,u,v)\to (A\symdif uu',u',v)]}{1-\pworm[(A,u,v)\to (A,u,v)]} & u\neq v,
\end{cases}
\\
\pwormprime[(A,u,u)\to (A,u,u)] &= \pworm[(A,u,u)\to (A,u,u)].
\end{split}
\label{rejection-free worm P}
\end{equation}
All other transitions occur with zero probability. In particular, no identity transitions are allowed from non-Eulerian states.
Corollary~\ref{rejection free corollary} immediately implies that $\pwormprime$ can be used to simulate $\phi_{G,n,x}$ for any $0 < n,\, x < \infty$.
We now proceed to show that in fact $\pwormprime$ remains valid even at $x=+\infty$.

\begin{remark}
  While the explicit conditioning \eqref{rejection-free worm P} is perhaps the simplest way to ensure ergodicity is retained as $x\to\infty$, there are variations of this idea that also work.
  Indeed, the algorithm presented in~\cite{ZhangGaroniDeng09} for the $n=1$ case was constructed in a slightly different way; the main consequence is that while configurations with $d_u(A)=d_v(A)=1$ were allowed
  in~\cite{ZhangGaroniDeng09}, they are forbidden as $x\to\infty$ in the algorithm we present here, as we shall now see.
\end{remark}

\subsection{Worm dynamics for fully-packed loops on bipartite cubic graphs}
\label{FPL worm algorithm section}
\begin{figure}[t]
  \caption{\label{distinct topologies} Possible topologies of the defect cluster $C_{uv}(A)$ when $d_u(A)=3$.}
  \begin{center}
    \psset{unit=0.4cm}
    \subfigure[Tadpole graph.]{\label{tadpole figure}
      \includegraphics{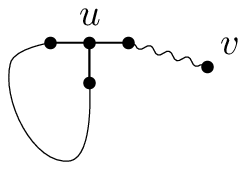}
    }
    \subfigure[Dumbbell graph.]{\label{dumbbell figure}
      \includegraphics{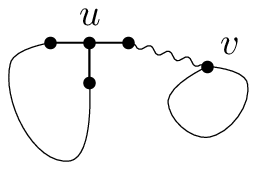}
    }
    \subfigure[Theta graph.]{\label{theta graph figure}
      \includegraphics{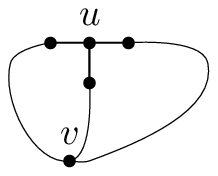}
    }
  \end{center}
\end{figure}

We now consider the $x\to\infty$ limit of $\pwormprime$.
We begin by noting that whenever $G$ has maximum degree $\Delta(G)\le3$ there are only a small number of possible topologies that the connected components of states in $\wormspace$ can have
(see Fig.~\ref{distinct topologies}).
\begin{proposition}
  \label{topological classification}
  Let $G$ be a finite graph with $\Delta(G)\le3$, let $(A,u,v)\in\mathcal{S}(G)$, and let $C_{uv}(A)$ be the component containing $u$ and $v$ in $(V,A)$.
  Then all components other than $C_{uv}(A)$ are isolated vertices or cycles, and we have the following classification of the possible topologies of the component $C_{uv}(A)$
  \begin{equation*}
    C_{uv}(A)=
    \begin{cases}
      \text{{\rm isolated vertex}}         & d_u(A)=d_v(A)=0\\
      \text{{\rm path}}                    & d_u(A)=d_v(A)=1\\
      \text{{\rm cycle}}                   & d_u(A)=d_v(A)=2\\
      \text{{\rm tadpole graph}}           & d_u(A)=3, d_v(A)=1$ \text{ or } $d_u(A)=1, d_v(A)=3\\
      \text{{\rm dumbbell or theta graph}} & d_u(A)=d_v(A)=3\\
    \end{cases}
  \end{equation*}
\end{proposition}
\noindent Proposition~\ref{topological classification} is intuitively obvious and its proof (which we omit) is straightforward.

For the remainder of this section we shall restrict attention to bipartite cubic graphs. On such graphs, the limit as $x\to\infty$ of $\pwormprime$ is now easily seen to be given by
\begin{multline}
  \lim_{x\to\infty }\, \pwormprime[(A,u,v)\to(A \symdif uu',u',v)]\\
  =
  \begin{cases}
    1/6      & u=v, uu'\not\in A,\\
    1/4      & d_u(A)=d_v(A)=1, uu'\not\in A,\\
    1/2      & d_u(A)=1, d_v(A)=3, uu'\not\in A,\\
    1/6      & C_{uv}(A) \text{ is a theta graph},\\
    n/2(n+2) & C_{uv}(A) \text{ is a dumbbell graph}, u\not\leftrightarrow u' \text{ in } (V,A\setminus uu'),\\
    1/2(n+2) & C_{uv}(A) \text{ is a dumbbell graph}, u    \leftrightarrow u' \text{ in } (V,A\setminus uu'),\\
  \end{cases}
  \label{P limit}
\end{multline}
and
\begin{equation}
\lim_{x\to\infty }\, \pwormprime[(A,u,u)\to(A,u,u)]=2/3.
\end{equation}
All other transitions occur with zero probability.

Now consider the subspace
\begin{equation}
\mathcal{R}(G) := \{(A,u,v)\in\wormspace \, : \, d_x(A) \neq 0 \text{ for all } x \text{ and } d_u(A)+d_v(A)\ge 4\}.
\label{FPL worm space}
\end{equation}
If $(A,u,v)\in\mathcal{R}(G)$, then either $|A|=|V|$ or $|A|=|V|+1$.
Furthermore, since \eqref{P limit} only allows the deletion of edges when $|A|=|V|+1$, it is clear that
$$
\lim_{x\to\infty }\, \pwormprime[(A,u,v)\to(A \symdif uu',u',v)] = 0
$$
whenever $(A,u,v)\in\mathcal{R}(G)$ and $(A\symdif uu',u',v)\not\in\mathcal{R}(G)$, so $\mathcal{R}$ is closed (and therefore recurrent).
The restriction of $\pworminftyprime$ to $\mathcal{R}(G)$ therefore defines a Markov chain on $\mathcal{R}(G)$.
We emphasize that the set of all bond configurations $A$ for which $(A,v,v)\in\mathcal{R}(G)$ corresponds precisely with $\fp$.
\begin{remark}
Since $G$ is a bipartite cubic graph, we know that $|A|=|V|$ for all $A\in\fp$.
It is therefore natural to consider $\lim_{x\to\infty }\, x^{-|V|}\,Z_{G,n,x}\,\piwormprime(A,u,v)$, where $Z_{G,n,x}$ is the appropriate normalization constant (partition function).
It is straightforward to verify that this limiting measure is supported on $\mathcal{R}(G)$ and is detailed balance with \eqref{P limit}.
\end{remark}
\begin{remark}
  The space $\mathcal{R}(G)$ given by \eqref{FPL worm space} is strictly smaller than the state space of the worm dynamics considered in~\cite{ZhangGaroniDeng09}.
  In particular, no states in which the defect cluster is a path are allowed in \eqref{FPL worm space}.
\end{remark}

Before proceeding further, it is useful to note that $\mathcal{R}(G)$ has the disjoint partition $\mathcal{R}(G)=\Eulerian\cup\Tadpole\cup\Theta\cup\Dumbbell$, where
\begin{equation}
\label{state space partition}
\begin{split}
\Eulerian &= \{(A,u,v)\in \mathcal{R}(G) \,: \, C_{uv}(A) \text{ is a cycle}\},\\
\Tadpole  &= \{(A,u,v)\in \mathcal{R}(G) \,: \, C_{uv}(A) \text{ is a tadpole graph}\},\\
\Theta    &= \{(A,u,v)\in \mathcal{R}(G) \,: \, C_{uv}(A) \text{ is a theta graph}\},\\
\Dumbbell &= \{(A,u,v)\in \mathcal{R}(G) \,: \, C_{uv}(A) \text{ is a dumbbell graph}\}.
\end{split}
\end{equation}
Now let us denote the restriction of \eqref{P limit} to $\mathcal{R}(G)$ by $\pworminftyprelim$.
From \eqref{P limit} we can see that the only transitions from $(A,u,u)\in\Eulerian$ which occur with non-zero probability are $(A,u,u)\to(A\cup uu',u',u)$ and $(A,u,u)\to(A\cup uu', u,u')$,
which both occur with the same probability $1/6$, and the identity transition $(A,u,u)\to (A,u,u)$.
We are therefore free to multiply the two transition probabilities $\pworminftyprelim[(A,u,u)\to(A\symdif uu',u',u)]=1/6$ from states $(A,u,u)\in\Eulerian$
by a constant factor $0<\gamma\le3$, provided that we also redefine the probabilities for identity transitions, so that we retain correctly-normalized row sums.
The only effect of such a modification is to multiply the stationary probabilities of all the states $(A,u,u)\in\Eulerian$ by the same constant $1/\gamma$.
Therefore, such modifications do not affect detailed balance.
If we now choose $\gamma=3$, then the identity transitions from $\Eulerian$ will occur with zero probability.
We thus obtain an entirely rejection-free Monte Carlo method.

Putting all these details together, let us now define the following transition matrix $\pworminfty$ on $\mathcal{R}(G)$
\begin{multline}
  \label{fpl worm transition matrix}
\pworminfty[(A,u,v)\to(A\triangle uu',u',v)] = \pworminfty[(A,v,u)\to(A\triangle uu',v,u')]\\
=
\begin{cases}
  1/2      & (A,u,v)\in \Eulerian\cup\Tadpole \text{ and $uu'\not\in A$},\\
  1/6      & (A,u,v)\in \Theta,\\
  n/2(n+2) & (A,u,v)\in \Dumbbell \text{ and $uu'$ is a bridge},\\
  1/2(n+2) & (A,u,v)\in \Dumbbell \text{ and $uu'$ is not a bridge}.
\end{cases}
\end{multline}
All other transitions are assigned zero probability; in particular, no identity transitions are allowed.
The transition matrix $\pworminfty$ corresponds to a very simple dynamics: if $|A|=|V|$ we must add one of the vacant edges incident to one of the defects; if $|A|=|V|+1$
must delete one of the occupied edges incident to one of the defects.
\begin{remark}
  Note that the only appearance of $n$ in \eqref{fpl worm transition matrix} occurs when $d_u(A)=d_v(A)=3$.
  Loosely, \eqref{fpl worm transition matrix} says that for transitions from degree 3 defects, the relative weight given by $\pworminfty$ of traversing a bridge is $n$,
  while the relative weight of traversing a non-bridge is $1$.
\end{remark}

It is now elementary to show that $\pworminfty$ is in detailed balance with the following distribution on $\mathcal{R}(G)$
\begin{equation}
\piworminfty(A,u,v)
=
\frac{n^{c(A)}}{Z_n}\,
\begin{cases}
  1        & (A,u,v)\in \Eulerian\cup\Tadpole,\\
  3/n        & (A,u,v)\in \Theta,\\
  (n+2)/n   & (A,u,v)\in \Dumbbell.
\end{cases}
\label{fpl worm stationary distribution}
\end{equation}
We prove in the next section that $\pworminfty$ is also ergodic.
It then follows from \eqref{fpl worm stationary distribution} and Lemma~\ref{sub-chain} that $\pworminfty$ defines a valid Markov-chain Monte Carlo algorithm to simulate the fully-packed loop model on
any bipartite cubic graph. We summarize this algorithm in Algorithm~\ref{fpl algorithm}.
\begin{algorithm}
  \begin{algthm}[FPL worm algorithm] $\,$
    \label{fpl algorithm}
    \begin{algorithmic}
      \LOOP
      \STATE Current state is $(A,u,v)$
      \IF{$u=v$}
      \STATE Choose the unique edge $uu'\not\in A$
      \STATE Perform, UAR, either $(A,u,u)\to(A\cup uu',u',u)$ or $(A,u,u)\to(A\cup uu',u,u')$
      \ELSIF{$u\neq v$}
      \IF{$d_u(A)=1$ or $d_v(A)=1$ (say $u$)}
      \STATE Choose, UAR, one of the 2 vacant edges incident to $u$ (say $uu'$)
      \STATE Make the transition $(A,u,v)\to(A\cup uu',u',v)$
      \ELSIF{$d_u(A)=d_v(A)=3$}
      \STATE Choose, UAR, one of the 2 defects (say $u$)
      \STATE Choose with probability $\mathbb{P}(u')$ one of the 3 neighbors of $u$ (say $u'$)
      \STATE Make the transition $(A,u,v)\to(A\setminus uu',u',v)$
      \ENDIF
      \ENDIF
      \ENDLOOP
    \end{algorithmic}
  \end{algthm}
\end{algorithm}

We note that the only place in which the topology for the loop configuration enters into Algorithm~\ref{fpl algorithm} is via the definition of the function $\mathbb{P}(u')$,
which from \eqref{fpl worm transition matrix} is given by
\begin{equation}
  \mathbb{P}(u') =
  \begin{cases}
  1/3      & (A,u,v)\in \Theta,\\
  n/(n+2) & (A,u,v)\in \Dumbbell \text{ and $uu'$ is a bridge},\\
  1/(n+2) & (A,u,v)\in \Dumbbell \text{ and $uu'$ is not a bridge}.
  \end{cases}
  \label{degree 3 weights}
\end{equation}
See Section~\ref{connectivity and colouring section} for a discussion of some possible implementations of the required topological queries.

\begin{remark}
  The $n\to0$ limit of Algorithm~\ref{fpl algorithm} is well defined and non-trivial.
  Taking the $n\to0$ limit of~\eqref{fpl worm transition matrix} or~\eqref{degree 3 weights} we see that
  the only change to Algorithm~\ref{fpl algorithm} when $n=0$ is that we are forbidden to delete bridges.
  If $G$ is hamiltonian, this suggests that the set of recurrent Eulerian states should be the set of all Hamiltonian cycles of $G$.
  Indeed, the $n\to0$ limit of~\eqref{fpl worm stationary distribution} shows that $\pi_{G,n=0}$ uniformly samples Hamiltonian cycles,
  and it is supported on the subset of $\mathcal{R}(G)$ in which the spanning subgraphs $(V,A)$ are connected.

  While it therefore seems plausible that the $n\to0$ limit of Algorithm~\ref{fpl algorithm} provides a valid Monte Carlo method for uniformly sampling Hamiltonian cycles,
  our general proof of the ergodicity of~\eqref{fpl worm transition matrix} breaks down at $n=0$.
  Although Algorithm~\ref{fpl algorithm} is very well suited to simulating fully-packed loops for $n>0$, it is not perhaps the most natural nor the most efficient dynamics for the special case of $n=0$,
  and so we have not attempted to prove its ergodicity in this limit.
  A more natural worm dynamics for simulating Hamiltonian cycles is presented in~\cite{Jacobsen08,OberdorfFergusonJacobsenKondev06}, applying earlier ideas from~\cite{Mansfield82}.
  We expect the specialized dynamics presented in~\cite{Jacobsen08,OberdorfFergusonJacobsenKondev06} is more efficient than the $n\to0$ limit of Algorithm~\ref{fpl algorithm}, and it has the added
  advantage that it simultaneously simulates both Hamiltonian paths and Hamiltonian cycles.
\end{remark}

\subsection{\texorpdfstring{Ergodicity of $\pworminfty$}{Ergodicity at infinite edge weight}}
\label{proofs}
We begin by making some brief comments on our notation.
Consider a Markov chain on a state space $\mathcal{S}$, with transition matrix $P$.
We say $s\in \mathcal{S}$ communicates with $s'\in \mathcal{S}$, and write $s\comm s'$, if the chain may ever visit state $s'$ with positive probability, having started in state $s$.
We say states $s$ and $s'$ intercommunicate, and write $s\intercomm s'$, if $s\comm s'$ and $s'\comm s$.
Using this notation, a set of states $\mathcal{A}\subseteq \mathcal{S}$ is ergodic iff $s\intercomm s'$ for all $s,s'\in \mathcal{A}$. Finally, we say $P$ is ergodic if $\mathcal{S}$ is ergodic under $P$.

\begin{proposition}
  \label{ergodicity proposition}
  If $G$ is a finite, connected, bipartite cubic graph and $n>0$, then $\pworminfty$ is ergodic.
\begin{proof}
Let $G$ be a finite, connected, bipartite cubic graph, let $B\in\fp$ be an arbitrary, but fixed, fully-packed subgraph, and let $z\in V$ be an arbitrary, but fixed, vertex.
We begin by proving that $(A,u,v)\comm (B,z,z)$ for every $(A,u,v)\in\mathcal{R}(G)$.

Suppose, then, that $(A,u,v)\in\mathcal{R}(G)$. We can generate a new state from $(A,u,v)$ via the map $f:\mathcal{R}(G)\to\mathcal{R}(G)$
with $f(A,u,v)$ defined by the following prescription:
\begin{algorithmic}
  \IF{$d_u(A)=1$}
  \STATE Choose an edge $uu'\in B$ with $uu'\not\in A$
  \RETURN $(A\cup uu',u',v)$
  \ELSIF{$d_v(A)=1$}
  \STATE Choose an edge $vv'\in B$ with $vv'\not\in A$
  \RETURN $(A\cup vv',u,v')$
  \ELSIF{$d_u(A)=2$ and $A\neq B$}
  \STATE Choose $ww'\in B$ with $ww'\not \in A$
  \RETURN $(A\cup ww',w',w)$
  \ELSIF{$d_u(A)=2$ and $A = B$}
  \RETURN $(B,z,z)$
  \ELSIF{$d_u(A)=3$}
  \STATE Choose the edge $uu'\not\in B$
  \RETURN $(A\setminus uu',u',v)$
  \ENDIF
\end{algorithmic}

The key observation to make is that $(A,u,v)\comm f(A,u,v)$ for every $(A,u,v)\in\mathcal{R}(G)$.
Indeed, if $d_u(A)\neq 2$ we simply have $\pworminfty[(A,u,v)\to f(A,u,v)]>0$.
Suppose instead that $d_u(A)=2$, which implies $u=v$. Lemma~\ref{eulerian defect move lemma} shows that $(A,u,u)\intercomm(A,w,w)$ for all $u,w\in V$,
so if $A=B$ then we clearly have $(A,u,v)=(B,u,u)\intercomm(B,z,z)=f(A,u,v)$. On the other hand, if $A\neq B$, then there must be at least one edge $ww'$ which is in $B$ but not in $A$.
Again, Lemma~\ref{eulerian defect move lemma} shows that $(A,u,u)\intercomm(A,w,w)$, and in addition we have
\begin{equation*}
    \pworminfty[(A,w,w)\to(A\cup ww',w',w)]>0,
\end{equation*}
so that $(A,w,w)\comm(A\cup ww',w',w)$, and consequently $(A,u,u)\comm(A\cup ww',w',w)$.
Therefore, we indeed have $(A,u,v)\comm f(A,u,v)$, and in fact $(A,u,v)\comm f^n(A,u,v)$ for any $n\in\mathbb{N}$, where $f^n=f\circ f\circ \dots \circ f$ denotes $n$-fold composition of $f$ with itself.

Now, whenever $A\neq B$, the state $f(A,u,v)$ has either one more occupied $B$-edge, or one less occupied non-$B$-edge, compared to $(A,u,v)$.
Therefore, since there are only a finite number of edges in $G$, if we start in any $(A,u,v)\in\mathcal{R}(G)$
and apply $f$ repeatedly, then we must eventually have $f^n(A,u,v)=(B,z,z)$, with $n$ necessarily {\em finite}.
It then immediately follows that $(A,u,v)\comm(B,z,z)$. Finally, the reversibility of $\pworminfty$ now implies that in fact $(A,u,v)\intercomm(B,z,z)$,
and since this holds for all $(A,u,v)\in\mathcal{R}(G)$ this implies $\pworminfty$ is ergodic.
\end{proof}
\end{proposition}

\begin{lemma}
  \label{eulerian defect move lemma}
  Let $G$ be a finite, connected, bipartite cubic graph. For every $A\in\fp$ and every pair $x,y\in V$ we have $(A,x,x)\intercomm(A,y,y)$ under $\pworminfty$.
  \begin{proof}
    We begin by showing that $(A,x,x)\comm (A,x',x')$ for all $x\in V$ and $x'\sim x$, where $x'\sim x$ denotes that $x'$ and $x$ are adjacent (i.e. they are neighbors).
    Firstly, note that if $xx'\not\in A$, then we simply have $\pworminfty[(A,x,x)\to(A\cup xx',x',x)]>0$ and $\pworminfty[(A\cup xx',x',x)\to(A,x',x')]>0$,
    which immediately implies $(A,x,x)\comm(A,x',x')$ in this case.

    Therefore, let us consider $x''\sim x$ with $xx''\in A$.
    Lemma~\ref{existence of worm paths lemma} guarantees that there exists an alternating path, $\mathcal{P}=z_1\,z_2\ldots z_{2k}$, such that $z_1=x$,
    $z_{2k}=x''$, $z_i\,z_{i+1}\not\in A$ for $i$ odd, and $z_i\,z_{i+1}\in A$ for $i$ even.
    The key observation is that it is always possible, via transitions of $\pworminfty$, to move one defect along such a path while leaving the other defect fixed, which can be seen as follows.
    Since $(A,z_1,z_1)\in \Eulerian$, we can always make the transition $(A,z_1,z_1)\to(A\cup z_1z_2, z_2,z_1)$ when $z_1z_2\not\in A$.
    Furthermore, since $|A|=|V|$ and $z_3\neq x$, it follows that $(A\cup z_1z_2,z_2,z_1)\in\Dumbbell\cup\Theta$ and $(A\cup z_1z_2\setminus z_2z_3,z_3,z_1)\in\Tadpole$.
    In general, if the position of the first defect is $z_i$ with $i$ even, then the corresponding state will be in $\Dumbbell\cup\Theta$, and the transition that
    moves the first defect from $z_i$ to $z_{i+1}$ by deleting the edge $z_iz_{i+1}$ will occur with strictly positive probability.
    Conversely, if the position of the first defect is $z_i$ with $i>1$ odd, then the corresponding state will be in $\Tadpole$, and since $z_iz_{i+1}\not\in A$, the transition that
    moves the first defect from $z_i$ to $z_{i+1}$ by adding the edge $z_iz_{i+1}$ will also occur with strictly positive probability.
    Consequently,
    $$
    (A,z_1,z_1) \comm \left(A\symdif E(\mathcal{P}),\,z_{2k},z_1\right)\,\in \Dumbbell\cup\Theta.
    $$

    Now we can move the second defect along $\mathcal{P}$, leaving the first defect fixed at $z_{2k}$. This has the effect of flipping each edge back to its original state, so we arrive at $(A,x'',x'')$.
    Indeed, an analogous argument to that above shows that
    \begin{equation*}
      (A\symdif E(\mathcal{P}),z_{2k},z_1)\comm(A\symdif E(\mathcal{P})\symdif E(\mathcal{P}),z_{2k},z_{2k}) = (A,x'',x'').
    \end{equation*}

    Therefore, $(A,x,x)\comm(A,x',x')$ for all $x'\sim x$. However, since this was proved for general $x$, it follows that precisely the same argument could be applied to show that
    $(A,x',x')\comm(A,x,x)$, and so we have $(A,x,x)\intercomm(A,x',x')$.
    Finally, since $G$ is connected, transitivity immediately implies that in fact $(A,x,x)\intercomm(A,y,y)$ for every $y\in V$.
  \end{proof}
\end{lemma}

\begin{remark}
  We remark that the dynamics of the worm algorithm along alternating paths discussed in Lemma~\ref{eulerian defect move lemma}, is very similar to the use of alternating paths by graph theorists in the
  construction of maximal matchings; see e.g.~\cite{Diestel05}.
\end{remark}
\bigskip

\begin{lemma}
  \label{existence of worm paths lemma}
  Let $G$ be a finite, connected, bipartite cubic graph.
  For every $A\in \fp$, every $x\in V$, and every $x'\sim x$, there exists a path $z_1\,z_2\ldots z_{2k}$ in $G$ such that $z_1=x$, $z_{2k}=x'$,
  $z_i\,z_{i+1}\not\in A$ for $i$ odd, and $z_i\,z_{i+1}\in A$ for $i$ even.
  \begin{proof}
    We begin by noting that any path $\mathcal{P}$ between $x$ and $x'\sim x$ must have odd length, because $\mathcal{P}+xx'$ is a cycle and $G$ is bipartite.

    Now, let $A\in \fp$ and $x\in V$, and suppose $xx'\not\in A$, $xx'',xx'''\in A$.
    The path $xx'$ is then trivially a path of the above form, with $k=1$, so let us focus on constructing a path from $x$ to $x''$.

    Let $z_1\,z_2\,\ldots\,z_{2k}\,z_1$ be a cycle in $(V,A)$.
    Since it contains an even number of edges, we can colour half of them blue, and half of them red, in such a way that each vertex $z_i$ is incident to precisely one blue edge and one red edge.
    For example, we can colour each edge $z_{i}\,z_{i+1}$ blue if $i$ is even and red if $i$ is odd. In this way the edges alternate red, blue, \ldots, red, blue as we traverse the cycle.
    Since the cycles in $(V,A)$ are vertex disjoint, such colourings can be performed independently for each cycle.

    After performing such a colouring, each vertex in $(V,A)$ is incident to precisely one red, one blue, and one vacant edge.
    If we now colour each vacant edge green, then we obtain a proper 3-edge-colouring of $G$.
    Suppose we now interpret the red edges as vacant, and the blue and green edges as occupied.
    Each vertex will again have degree 2 (one blue edge plus one green edge), so this procedure generates a new bond configuration $A_{\rm red}\in\fp$.
    Furthermore, since each vertex is incident to precisely one green edge and one blue edge, each cycle $z_1\,z_2\,\ldots\,z_{2k}\,z_1$ in $A_{\rm red}$ must be such that
    the edges alternate green, blue, \ldots, green, blue as we traverse the cycle.

    Now, fix $A\in \fp$ and $x\in V$, and suppose $xx'\not\in A$, $xx'',xx'''\in A$. Suppose we perform a colouring of $A$, as described above, in which the edge $xx''$ is blue.
    It then follows that there is a cycle $z_1\,z_2\,\ldots z_{2k}\,z_1$ in $(V,A_{\rm red})$ in which $z_1=x$, $z_2=x'$ and $z_{2k}=x''$. But this defines a path $z_1\,z_2\,\ldots\,z_{2k}$
    in $G$ in which the edges $z_iz_{i+1}$ are green (vacant in $(V,A)$) when $i$ is odd, and blue (occupied in $(V,A)$) when $i$ is even.
    We have therefore showed that there exists an alternating path of the required form between $x$ and $x''$.
    A similar argument can obviously be used to construct the required path between $x$ and $x'''$; in fact the same colouring can be used as for the $xx''$ path, provided we interpret the red and green
    edges as occupied, and the blue edges as vacant.
  \end{proof}
\end{lemma}

\subsection{Connectivity-checking and the colouring method}
\label{connectivity and colouring section}
An important practical matter when implementing the algorithms we have presented so far, is the need, when $n\neq 1$,
to perform a non-local query to determine if the cyclomatic number changes when an update is performed.
Consider a spanning subgraph $(V,A)\subseteq G$ of a graph $G=(V,E)$.
Since the number of components, $k(A)$, is related to the cyclomatic number, $c(A)$, by $k(A) = |V| - |A| + c(A)$,
the task of determining whether an edge-update changes the cyclomatic number is equivalent to determining whether it
changes the number of connected components. The latter question can be answered by known dynamic connectivity-checking algorithms~\cite{HolmDeLichtenbergThorup01}, which take polylogarithmic amortized time.
A much simpler approach, which runs in polynomial time, but with a (known) small exponent is simultaneous breadth-first search~\cite{DengZhangGaroniSokalSportiello10}.
We used the latter approach in the simulations presented in Section~\ref{numerical section}.
In Sections~\ref{colouring method section} and~\ref{loop model colouring method section} we discuss a different approach, the {\em colouring method},
which avoids altogether the need for such global queries, at least when $n>1$.
Before discussing the colouring method, however, we make some remarks regarding the practical implementation of connectivity queries.

To illustrate, we will consider Algorithm~\ref{fpl algorithm} with $n>1$.
Algorithm~\ref{fpl algorithm} states that connectivity queries are necessary only when $(A,u,v)\in\Theta\cup\Dumbbell$.
In practice, however, even in this case one does not usually need to perform such queries.
Suppose we assign a fixed (but arbitrary) ordered labeling to $V$, so $V=\{v_1,v_2,\ldots\}$, and let $r\in[0,1]$ be a uniformly-distributed random number.
For notational convenience we set $p=1/(n+2)$ and $\epsilon=1/3-p>0$.
We can implement the $d_u(A)=d_v(A)=3$ block in Algorithm~\ref{fpl algorithm} as follows.
Denote the neighbors of $u$ by $u_j$ with $j=1,2,3$, such that $u_1$ has the smallest label, and $u_3$ the largest.
If $r\in[(j-1)\,p,j\,p]$ then we can simply choose $u_j$ without knowing the topology of the defect cluster.
If instead $r>3\,p$, then we need to determine whether or not any of the edges $uu_j$ are bridges; there can be at most one.
If $uu_j$ is a bridge then we choose $u_j$, otherwise if none of the $uu_j$ are bridges then we choose $uu_j$ iff $r\in[3\,p + (j - 1)\,\epsilon, 3\,p + j\,\epsilon]$.
An analogous trick can be employed when $n<1$.

The computational burden imposed by these connectivity queries is of greater concern when $x<\infty$ than when $x=\infty$.
At $x=+\infty$, connectivity-checking is only required when $d_u(A)=d_v(A)=3$.
By contrast, when $x<\infty$ it is in principle always necessary unless both defects are isolated, except where avoided by a trick of the type described above.
For this reason, one might expect that the colouring method (whose raison d'\^etre is to avoid connectivity queries) would be more advantageous when $x<\infty$.
Our simulations suggest that this is indeed the case.
In fact, while the colouring method was significantly more efficient on the critical branch, it was significantly {\em less} efficient than the connectivity-checking version when $x=\infty$.

\subsubsection{The colouring method}
\label{colouring method section}
Consider a finite graph $G=(V,E)$.
The {\em colouring method}~\cite{ChayesMachta98,DengGaroniMachtaOssolaPolinSokal07,DengGaroniGuoBloeteSokal07} is a general methodology for simulating models of the form
\begin{equation}
\phi_{G,W}(A)\propto
\prod_{C \in K(A)} W(C), \qquad A\subseteq E,
\label{CW measure}
\end{equation}
where $K(A)$ denotes the set of all connected components of $(V,A)$, and $W$ is an arbitrary map that associates a nonnegative weight to every connected subgraph of $G$.
Many lattice models in statistical mechanics can be expressed in the form~\eqref{CW measure}. If we set $W(C)=q\,v^{|E(C)|}$ for all $C$, then we recover the standard random-cluster model of
Fortuin-Kasteleyn~\cite{Grimmett06}. If, instead, we set
\begin{equation}
W(C) =
\begin{cases}
  1, & \text{if $C$ is an isolated vertex},\\
  n\,x^{|E(C)|}, & \text{if $C$ is a cycle},\\
  0, & \text{otherwise,}\\
\end{cases}
\label{c(A) colour weight}
\end{equation}
and $G$ has maximum degree $\Delta(G)\le3$, then $\phi_{G,W}=\phi_{G,n,x}$ and we recover the loop model~\eqref{loop measure}.

The key step in applying the colouring method is to choose an appropriate nonnegative weight function $\activeW < W$ for which we have a transition matrix $P_{G,\activeW}$ to simulate $\phi_{G,\activeW}$.
In practice, {\em appropriate} means that $P_{G,\activeW}$ is more efficient/convenient to implement than any algorithm we have at hand for $\phi_{G,W}$.
We shall return to this point in Section~\ref{loop model colouring method section}.
Given $\activeW$ and $P_{G,\activeW}$, the colouring method simulates $\phi_{G,W}$ by an algorithm which, at each step,
updates the bonds on a suitably-chosen random subgraph $H\subseteq G$ using $P_{H,\activeW}$, while leaving all other bonds fixed.
  \begin{algthm}[colouring method] $\,$
    \label{colouring algorithm}
    \begin{algorithmic}
      \LOOP
      \STATE Current state is $A\subseteq E$
      \STATE Independently colour each $C\in K(A)$ red with probability $\activeW(C)/W(C)$ and blue otherwise
      \STATE Identify the active subgraph $G_{\active}=G[V_{\active}]$
      \STATE Choose a new $A_{\rm red}'$ via $P_{G_{\rm red},\activeW}[A_{\active}\to A_{\active}']$
      \STATE New state is $A'=A_{\rm red}'\cup A_{\rm blue}$
      \ENDLOOP
    \end{algorithmic}
  \end{algthm}
\noindent By independently colouring each cluster in $K(A)$ we obtain a random 2-colouring of the vertices $\colour\in\{\active,\inactive\}^V$ for which each edge in $A$ has both its endpoints coloured
the {\em same} colour. We consider the subgraph induced by the red vertices $G_{\active}(\colour)=G[V_{\active}(\colour)]$ as {\em active} and that induced by the blue vertices as {\em frozen}.
The set $A_{\active}= A \cap E(G_{\active})$ is the subset of all edges in $A$ for which both endpoints lie in $V_{\active}$, and similarly $A_{\inactive}=A\cap E(G_{\inactive})$.

In Section~\ref{loop model colouring method section} we specialize Algorithm~\ref{colouring algorithm} to construct a worm algorithm for the loop model $\phi_{G,n,x}$,
which we present in Algorithm~\ref{worm colouring algorithm}.
We conclude the current section by providing a more precise statement of Algorithm~\ref{colouring algorithm} in terms of transition matrices.
To this end, let us set $\mathbf{W}=(W_{\active},W_{\inactive})=(\activeW,W-\activeW)$ and introduce the following joint measure of colours and bonds
\begin{equation}
  \jointmeasure(\edgeset,\colour)
  \propto \Delta(\edgeset,\colour) \, \prod_{C\in K(\edgeset)} \, W_{\sigma(C)}(C),
  \label{joint measure}
\end{equation}
where $\sigma(C)$ denotes the colour of the vertices in cluster $C\in K(\edgeset)$,
and $\Delta(A,\colour)$ is the indicator for the event $\{ (A,\colour): \sigma_i=\sigma_j \text{ for all } ij\in A\}$.
The transition matrix of the colouring method provides a Monte Carlo algorithm to simulate the joint measure \eqref{joint measure}.
Since the marginal measure $\sum_{\colour}\jointmeasure(\cdot,\colour)$ on $\{A\subseteq E\}$ is simply $\phi_{G,W}$,
it then follows immediately that the colouring method in fact provides a Monte Carlo method for $\phi_{G,W}$.
Proposition~\ref{colouring method} provides a precise definition as well as a justification of the colouring-method transition matrix.
Recall that the support of $\jointmeasure$ is, by definition, $\supp(\jointmeasure):=\{(\edgeset,\colour): \jointmeasure(\edgeset,\colour)>0\}$.
In a slight abuse of notation, we also write $\colour\in\supp(\jointmeasure)$ whenever there exists $(A,\colour)\in\supp(\jointmeasure)$.
\begin{proposition}[colouring Method]
  Consider $\activeW<W$ with $\supp(\activeW)=\supp(W)$, and for each $\colour\in\supp(\jointmeasure)$ let $P_{G_{\active}(\colour),\activeW}$ be a transition matrix with state space
  $\{A_{\active}\subseteq E(G_{\active}(\colour))\}$ and stationary distribution $\phi_{G_{\active}(\colour),\activeW}$, and suppose
  $P_{G,\activeW}$ is ergodic on $\supp(\phi_{G,\activeW})$.
  If we define transition matrices $\pcolour$ and $\pbond$ on $\supp(\jointmeasure)$ by
  \begin{align*}
    \pcolour[(A,\colour) \to (A',\colour')] &= \delta_{A,A'}\Delta(A,\colour')\prod_{C\in K(A)}\frac{W_{\colour'(C)}(C)}{W(C)},
    \\
    \pbond[(A,\colour) \to (A',\colour')]
    &=
    \delta_{\colour,\colour'}\,\Delta(A',\colour)\,\delta_{A_{\inactive},A_{\inactive} '}\, P_{G_{\active}(\colour),\activeW}[A_{\active} \to A_{\active}'],
  \end{align*}
  then $P := \pcolour\,\pbond$ is ergodic and has stationary distribution $\jointmeasure$.
  \label{colouring method}
  \begin{proof}
    We simply sketch the proof; see~\cite{DengGaroniMachtaOssolaPolinSokal07,DengGaroniGuoBloeteSokal07} for further details.
    Ergodicity follows by noting that there is a positive probability of consecutively colouring the whole graph red an arbitrary number of times, and then relying on the ergodicity of $P_{G,\activeW}$.
    Stationarity of $\jointmeasure$ follows by observing that $\jointmeasure$ is in fact stationary with respect to both $\pcolour$ and $\pbond$
    separately, and these latter two facts can be easily verified by noting that $\pcolour[(\edgeset,\colour)\to(\edgeset',\colour')]=\delta_{\edgeset,\edgeset'}\,\jointmeasure(\colour'|\edgeset)$, and
    \begin{equation*}
      \jointmeasure(\edgeset\,|\,\colour)=\Delta(\edgeset,\colour)\,\phi_{G_{\active}(\colour),\activeW}(A_{\active})\,\phi_{G_{\inactive}(\colour),\activeW}(A_{\inactive}).
    \end{equation*}
  \end{proof}
\end{proposition}

\subsubsection{Applying the colouring method to worm algorithms for the loop model}
\label{loop model colouring method section}
Since we do not need to perform connectivity checks when $n=1$, we consider worm updates for the $n=1$ model to be {\em convenient}, and we know from experience~\cite{DengGaroniSokal07c,ZhangGaroniDeng09} that
they are also efficient. When $n>1$ it is therefore natural to choose
\begin{equation}
  \activeW(C)=
  \begin{cases}
    x^{|E(C)|}, & \text{if $C$ is Eulerian},\\
    0, & \text{otherwise.}\\
  \end{cases}
\end{equation}
The resulting algorithm proceeds as follows:
\begin{algthm}[coloured worm algorithm] $\,$
  \label{worm colouring algorithm}
  \begin{algorithmic}
    \LOOP
    \STATE Current state is $A$
    \STATE Colour each isolated vertex red
    \STATE Independently colour each loop red with probability $1/n$
    \STATE Identify $G_{\rm red}$
    \STATE Choose, uniformly at random, $v\in V(G_{\rm red})$
    \STATE Use $n=1$ worm updates on $G_{\active}$ to make a transition $(A_{\rm red},v,v)\to(A_{\rm red}',v',v')$
    \STATE New state is $A'=A_{\rm red}'\cup A_{\rm blue}$
    \ENDLOOP
  \end{algorithmic}
\end{algthm}

Finally, let us consider the $n=1$ worm updates in a little more detail. Let $H\subseteq G$ and consider the following transition matrix on $\mathcal{C}(H)$
\begin{equation}
  \label{pwormloop}
  \widetilde{P}_{H,x}[A\to A'] := \frac{1}{V}\,\sum_{v,v'\in V}\,\overline{P}_{H,n=1,x}[(A,v,v)\to(A',v',v')],
\end{equation}
where $\pwormbar$ is the restriction of \eqref{worm P} to the Eulerian subspace \eqref{eulerian subspace}.
$\widetilde{P}_{H,x}[A\to A']$ is the probability that, starting in $A\in\mathcal{C}(H)$, we pick, uniformly at random, a location for the defects, $v$,
then perform worm updates from $(A,v,v)$ until we arrive at
a new $A'\in\mathcal{C}(H)$, regardless of the new location of the defects. It is clear that the row sums of $\widetilde{P}_{H,x}$ are correctly normalized, so that it defines a stochastic matrix, and
it is also clear that $\widetilde{P}_{H,x}$ is in detailed balance with $\phi_{H,n=1,x}$.
In addition, since $\pwormbar$ is ergodic on any $G$ when $x<\infty$, it follows immediately that $\widetilde{P}_{H,x}$ is ergodic for all $x<\infty$.
The transition matrices $P_{G_{\active}(\colour),\activeW}$ required in Proposition~\ref{colouring method} are then chosen to be $\widetilde{P}_{G_{\active}(\colour),x}$.
Analogous transition matrices can obviously be constructed from \eqref{rejection-free worm P} and \eqref{fpl worm transition matrix}.
For the fully-packed case, $\widetilde{P}_{H,x=+\infty}$ is not necessarily ergodic on all possible subgraphs, however Proposition~\ref{colouring method} only requires that it be ergodic for $H=G$, which
is guaranteed by Proposition~\ref{ergodicity proposition}.

\begin{remark}
  Let us return to the case of the genuinely $n$-dependent connectivity-checking versions of the worm dynamics,
  as discussed in Sections~\ref{standard worm}, \ref{rejection-free section}, and~\ref{FPL worm algorithm section}.
  We note that the right-hand side of~\eqref{pwormloop}, with $n$ left arbitrary rather than fixed to $1$, defines a perfectly valid alternative worm algorithm, in which an additional step is added:
  whenever the defects collide, uniformly at random choose a new vertex to move them both to. In particular, for the FPL model, if one were to add such a move then ergodicity could be proved
  without recourse to Lemmas~\ref{eulerian defect move lemma} and~\ref{existence of worm paths lemma}. However, we find empirically that there is no practical advantage to adding these moves, and we did
  not use them in our simulations (except, of course, when using the colouring method).
\end{remark}

\section{Numerical results}
\label{numerical section}
We simulated the loop model \eqref{loop measure} on an $L \times L$ honeycomb lattice with periodic boundary conditions, using the algorithms described in Section~\ref{Worm algorithms}.
In particular, we used the genuinely $n$-dependent algorithms described in Sections~\ref{rejection-free section} and~\ref{FPL worm algorithm section}, as well as the colouring
algorithms described in Section~\ref{loop model colouring method section}.
We shall refer to these two distinct versions as the {\em connectivity-checking} and {\em colouring} versions, respectively.

We considered both the cases $n\le 2$ and $n>2$.
The questions studied differed substantially in these two cases, since the exact phase diagram is known when $n\le2$, while no exact results are known at all for $n>2$.
For each choice of $n$, and each possible branch (when $n\le2$) we simulated at least seven (and up to eleven) different choices $L$, in the range $12\le L\le L_{\max}$.
The values of $L_{\max}$ used depended on the choices of $x$ and $n$. They are summarized in Table~\ref{L values}.
\begin{table}[htb]
  \caption{\label{L values} Summary of different values of $L_{\max}$ used in each simulation.
  }
  \medskip
  \centering
      {\footnotesize
	\begin{tabular}{|c|cccc|ccc|cccccc|cc|}
	  \hline
	  \multicolumn{1}{|c}{} & \multicolumn{4}{c|}{Critical} & \multicolumn{3}{c|}{Densely-packed} & \multicolumn{6}{c|}{Fully-packed} & \multicolumn{2}{c|}{$n>2$} \\
	  \hline
	  $n$         & 0.5 & 1.0 & 1.5 & 2.0      & 0.5 & 1.0 & 1.5 & 0.1 & 1.0 & 1.25 & 1.5 & 1.75 & 2.0&  3 & 10 \\
	  $L_{\max}$  & 240 & 360 & 240 & 240      & 120 & 240 & 240 & 120 & 240 & 240  & 240 & 240  & 240 & 120 & 120 \\
	  \hline
	\end{tabular}
      }
\end{table}

For $n>2$, we simulated at $n=3$ and $10$, with the aim of verifying the existence of a phase transition, the nature of its universality class, and obtaining accurate estimates of the critical points.
For $n\le 2$, we simulated on both branches of \eqref{integrable curve}, as well as at $x=+\infty$.
We focused on identifying the scaling exponents of five distinct classes of observables, characterizing loop lengths, face sizes, the magnetization of dual Ising spin configurations
and its staggered analogue, and the return time to the Eulerian subspace.
Although the latter observable is, by construction, defined on the full worm space $\mathcal{S}(G)$, rather than on the loop state space $\mathcal{C}(G)$,
we shall see that it appears to have a deep connection with the loop model itself.

\subsection{Observables measured}
We measured the following observables in our simulations.
All observables were measured only when the defects coincided, with the exception of the return time, $\mathcal{T}$, which is defined on the full worm chain.
\begin{itemize}
\item The number of loops $\mathcal{N}_l(A)=c(A)$.
\item The number of bonds $\mathcal{N}_b(A)=|A|$.

\noindent Note that on the fully-packed branch we trivially have $\mathcal{N}_b=|V| = 2L^2$, so we only measured this quantity on the critical and densely-packed branches.
\item The length of the largest loop $\mathcal{L}_1$
\item The mean-square loop length
\begin{equation}
\mathcal{L}_2 := L^{-2}\,\sum_{l} |l|^2
\label{def_quantity_l2}
\end{equation}
where the sum is over all loops $l$.
\item The size of the largest face $\mathcal{G}_1$
\item The mean-square face size
\begin{equation}
\mathcal{G}_2 := L^{-2}\,\sum_{f} |f|^2
\end{equation}
where the sum is over all faces $f$.
Every loop configuration $A$ on the honeycomb lattice can be decomposed into a number of faces, each consisting of a collection of
elementary hexagons, such that every pair of neighboring elementary hexagons which share an unoccupied edge in $A$ belong to the same face.
The size $|f|$ of face $f$ is then simply the number of elementary hexagons which it contains.
\item The dual Ising magnetization $\mathcal{M}$.

\noindent This is measured by assigning an Ising configuration to the dual triangular lattice in such a way that the loops on the honeycomb lattice form
  the domain boundaries of the Ising spin configuration. Such Ising configurations can be defined in an unambiguous manner whenever the loop configuration winds the torus an even number of times,
  and we therefore only measured this observable when the loop configuration was in this subspace of the cycle space.
  In such cases the spin configuration is unique (up to a global spin flip $\sigma\mapsto-\sigma$).
\item The sublattice dual Ising magnetization $\mathcal{M}_{i}$, for $i=1,2,3$.

\noindent Since the triangular lattice is tripartite, we can independently consider the Ising magnetization on each of its three sublattices.
\item The return time $\mathcal{T}$ to the Eulerian subspace $\mathcal{C}(G)\times V$.
\end{itemize}
\bigskip
From these observables we estimated the following quantities:
\begin{itemize}
\item The loop-number density $n_l:= L^{-2} \langle \mathcal{N}_l\rangle$
\item The loop-number fluctuation $C_l := L^{-2} \text{var}(\mathcal{N}_l)$
\item The bond-number density on the critical and densely-packed branches $n_b:=L^{-2} \langle \mathcal{N}_b\rangle$
\item The bond-number fluctuation on the critical and densely-packed branches $C_b := L^{-2} \text{var}(\mathcal{N}_b)$
\item The expectations $\<\mathcal{L}_1\>$ and $\<\mathcal{L}_2\>$
\item The expectations $\<\mathcal{G}_1\>$ and $\<\mathcal{G}_2\>$
\item The dual Ising susceptibility $\chi_{\rm Ising}=L^{-2}\,\langle \mathcal{M}_{\rm Ising}^2\rangle$
\item The second and fourth moments, $\<\mathcal{M}_{\rm stag}^2\>$ and $\<\mathcal{M}_{\rm stag}^4\>$, of the the staggered dual Ising magnetization, $\mathcal{M}_{\rm stag}$, defined by
  $$
  \mathcal{M}_{\rm stag}^2 := (\mathcal{M}_1-\mathcal{M}_2)^2 + (\mathcal{M}_2-\mathcal{M}_3)^2 + (\mathcal{M}_3-\mathcal{M}_1)^2
  $$
\item The staggered susceptibility $\chi_{\rm stag} = L^{-2}\,\<\mathcal{M}_{\rm stag}^2\>$
\item The dimensionless ratio
  $$
  Q_s=\frac{\<\mathcal{M}_{\rm stag}^2\>^2}{\<\mathcal{M}_{\rm stag}^4\>}
  $$
\item The mean return time to the Eulerian subspace $\<\mathcal{T}\>$. We also estimated the higher-order moments $\<\mathcal{T}^k\>$, for $k=2,3,4$, and the distribution $\mathbb{P}(\mathcal{T}=t)$.
\end{itemize}
\bigskip

For each quantity $Y = n_l$, $n_b$, $C_l$, $C_b$, $\<\mathcal{L}_1\>$, $\<\mathcal{L}_2\>$, $\<\mathcal{G}_1\>$, $\<\mathcal{G}_2\>$, $\chi_{\rm Ising}$, $\chi_{\rm stag}$, $Q_s$
and $\<\mathcal{T}^k\>$ we performed a least-squares fit of our Monte Carlo data to the finite-size scaling (FSS) ansatz
\begin{equation}
  \begin{split}
    Y(\beta,L)
    =
    c_0 + c_1(\beta-\beta_c) + \dots + L^{x_{Y}}\left[\right. a_0 &+ a_1 (\beta-\beta_c)L^{y_t} + a_2 (\beta-\beta_c)^2 L^{2y_t}+\dots
      \\
      &\left.+ \, b_1 \, L^{-\omega_1}+b_2\,L^{-\omega_2} + \dots\right].
  \end{split}
  \label{FSS ansatz}
\end{equation}
Here $y_t$ is the leading thermal exponent.
The $a_i$ are coefficients of the FSS variable $(\beta-\beta_c)L^{y_t}$, the $b_i$ are the coefficients of the corrections-to-scaling terms, and the $c_i$ are the coefficients of analytic terms.
There are also cross-terms involving products of terms arising from each of these three sources. The choice of which terms to include in the fit for a given choice of observable varied from case to case,
and involved a certain amount of trial and error. The exponent $x_{Y}$ in \eqref{FSS ansatz} is a generic label for whatever the dominant exponent happens to be for the quantity $Y$.
In particular, if $Y$ happens to be dimensionless, such as $Q_s$, then we have $x_{Y}=0$ identically, and in this case all the $c_i$ are identically zero.

As a precaution against corrections to scaling, we imposed a lower cutoff $L\ge L_{\text{min}}$ on the data points admitted to the fit,
and we studied systematically the effects on the fit of varying the value of $L_{\text{min}}$.
We used the Levenberg-Marquardt algorithm to perform the fits.

\subsection{\texorpdfstring{Fits for $n\le 2$}{n <= 2}}
The results of the fits for $n\le 2$ are presented in Tables~\ref{critical branch exponents table}, \ref{DPL branch exponents table} and~\ref{FPL branch exponents table}.
We discuss these results observable by observable in the following sections.
\begin{table}[htb]
  \caption{\label{critical branch exponents table}Critical exponents for the critical branch.
    The exponent $X_{\rm worm}$ denotes the scaling dimension of $\<\mathcal{T}\>$ for the connectivity-checking version of the worm dynamics.
  }
  \medskip
  \centering
      {\footnotesize
	\begin{tabular}{|l|lllllllllll|}
	  \hline
	  $n$   &$X_{\rm loop}$ &$X_{\rm hull}$ &$X_{\rm face}$ &$\XhPotts$   &$X_{\rm Ising}$ &$\XtPotts$ &$X_{\rm energy}$ &$X_{\rm stag}$ &$\XtSubPotts$  & $X_{\rm worm}$ &$\XhOn$  \\
	  \hline
	  $0.5$ &$0.648(2)$     &$0.6478$       &$0.0258(4)$    &$0.0261$     &$0.0567(2)$     &$0.0567$   &$0.816(3)$      &$0.79(2)$      &$0.8178$        & $0.1157(3)$    &$0.1154$ \\
	  $1.0$ &$0.623(2)$     &$0.6250$       &$0.0518(4)$    &$0.0521$     &$0.1251(2)$     &$0.1250$   &$0.998(3)$      &$0.97(2)$      &$1.0000$        & $0.1250(2)$    &$0.1250$ \\
	  $1.5$ &$0.592(2)$     &$0.5935$       &$0.0800(4)$    &$0.0801$     &$0.2194(2)$     &$0.2195$   &$1.254(5)$      &$1.28(6)$      &$1.2519$        & $0.1322(2)$    &$0.1322$ \\
	  $2.0$ &$0.5000(2)$    &$0.5000$       &$0.1249(3)$    &$0.1250$     &$0.4998(3)$     &$0.5000$   &$1.996(7)$      &\---           &$2.0000$        & $0.1251(5)$    &$0.1250$ \\
	  \hline
	\end{tabular}
      }
\end{table}

\begin{table}[htb]
  \caption{\label{DPL branch exponents table}Critical exponents for the densely-packed loop branch.
    The exponent $X_{\rm worm}$ denotes the scaling dimension of $\<\mathcal{T}\>$ for the connectivity-checking version of the worm dynamics.
    No exponents are reported for $\chi_{\rm stag}$ since it was found to be constant.
  }
  \medskip
  \centering
      {\footnotesize
	\begin{tabular}{|l|llllllll|}
	  \hline
	  $n$   &  $X_{\rm loop}$  & $X_{\rm hull}$ & $X_{\rm face}$ & $\XhPotts$   & $X_{\rm Ising}$ & $\XtPotts$   & $X_{\rm worm}$          & $\XhOn$             \\
	  \hline
	  $0.5$ &  $0.139(2)$      &  $0.1386$      & $0.0640(4)$    & $0.0637$     &  $1.58(1)$      & $1.5843$     & $-0.0788(4)$            & $-0.0791$           \\
	  $1.0$ &  $0.2500(1)$     &  $0.2500$      & $0.1044(4)$    & $0.1042$     &  \---           & \---         & $\phantom{-}0.0000(1)$  & $\phantom{-}0.0000$ \\
	  $1.5$ &  $0.3508(3)$     &  $0.3506$      & $0.1282(3)$    & $0.1280$     &  $0.947(3)$     & $0.9482$     & $\phantom{-}0.0620(3)$  & $\phantom{-}0.0619$ \\
	  \hline
	\end{tabular}
      }
\end{table}

\begin{table}[htb]
  \centering
  \caption{\label{FPL branch exponents table}Critical exponents on the fully-packed branch.
    The exponent $X_{\rm worm}$ denotes the scaling dimension of $\<\mathcal{T}\>$ for the connectivity-checking version of the worm dynamics.
  }
  \medskip
      {\footnotesize
	\begin{tabular}{|l|llllllllll|}
	  \hline
	  $n$   & $X_{\rm loop}$  & $X_{\rm hull}$  & $X_{\rm face}$ & $\XhPotts$    & $X_{\rm Ising}$      & $\XtPotts$        & $X_{\rm stag}$ & $2/3g$    & $X_{\rm worm}$ & $\XhOn$   \\
	  \hline
 	  $0.1$ & $0.031(3)$      & $0.0309$        & $0.0154(4)$    & $0.0152$      & $ 1.92(6) $          & $ 1.9074$         & $0.322(5)$     & $0.3231$  & $0.032(4)$     & $0.0309$  \\
 	  $1.0$ & $0.2500(2)$     & $0.2500$        & $0.1042(3)$    & $0.1042$      & \---                 & $ 1.2500$         & $0.2499(1)$    & $0.2500$  & $0.2498(4)$    & $0.2500$  \\
 	  $1.25$& $0.3005(2)$     & $0.3006$        & $0.1181(2)$    & $0.1180$      & $ 1.10(3) $          & $ 1.0982$         & $0.2333(3)$    & $0.2331$  & $0.3005(3)$    & $0.3006$  \\
 	  $1.5$ & $0.3505(4)$     & $0.3506$        & $0.1276(4)$    & $0.1280$      & $ 0.93(2) $          & $ 0.9482$         & $0.2165(3)$    & $0.2165$  & $0.3507(3)$    & $0.3506$  \\
 	  $1.75$& $0.4039(3)$     & $0.4042$        & $0.1336(3)$    & $0.1335$      & $ 0.79(1) $          & $ 0.7875$         & $0.1986(4)$    & $0.1986$  & $0.4043(5)$    & $0.4042$  \\
 	  $2.0$ & $0.482(3)$      & $0.5000$        & $0.1345(3)$    & $0.1250$      & $ 0.57(2) $          & $ 0.5000$         & $0.175(3)$     & $0.1667$  & $0.480(3)$     & $0.5000$  \\
	  \hline
	\end{tabular}
      }
\end{table}

\subsubsection{\texorpdfstring{Scaling of energy-like quantities}{Scaling of the energy-like quantities}}
Standard finite-size scaling arguments predict that on the critical branch
\begin{align}
  n_b &\sim a + b\,L^{y_{\rm energy}-2}, \label{eq_fit_nb} \\
  C_b &\sim a + b\,L^{2\,y_{\rm energy}-2}, \label{eq_fit_cb}
\end{align}
where $y_{\rm energy} = 2 - X_{\rm energy}$ is the fractal dimension characterizing the energy-like quantities, and $X_{\rm energy}$ is the corresponding scaling dimension.
It was observed in~\cite{DengGaroniGuoBloeteSokal07} that for the critical loop model with $1<n<2$, we have $X_{\rm energy}=\XtSubPotts$
where $\XtSubPotts$ is the second thermal scaling dimension of the $q$-state Potts model with $q=n^2$, whose expression in terms of the Coulomb gas coupling $g$ is known~\cite{Nienhuis84} to be
\begin{equation}
  \XtSubPotts= -2 + 16/g.
  \label{subpotts_exponent}
\end{equation}
For $n=1$, one has $\XtSubPotts=1$ and $2\,y_{\rm energy}-2=0$, and Eq.~(\ref{eq_fit_cb}) is replaced by $C_b \sim a + b\, \ln L$.
Table~\ref{critical branch exponents table} shows the numerical results, which confirms Eq.~(\ref{subpotts_exponent}).

The behaviour of the loop-number density  $n_l$ and its fluctuation $C_l$ is also described by Eqs.~(\ref{eq_fit_nb}) and~(\ref{eq_fit_cb}). This is perhaps unsurprising in light of the Euler relation $c(A) = |A| - |V| + k(A)$.

On any graph of maximum degree 3, the observable $\mathcal{N}_b/\mathcal{N}_l$ is simply the arithmetic mean of the loop lengths in a given configuration.
This quantity was studied in detail in~\cite{JacobsenVannimenus99}, and in particular it was found that its expectation $\<\mathcal{N}_b/\mathcal{N}_l\>$ tends
to a constant in the thermodynamic limit.

On the critical branch, the fitting results for the constants,  $n_{b0}$, $C_{b0}$, $n_{l0}$,  $C_{l0}$, and  $B=n \; n_{b0}/n_{l0}$, are shown in
Table \ref{Constants table}. Our estimate $B(n=2) = 38.832(2)$ agrees well with  $38.834 (2)$ in Ref.~\cite{JacobsenVannimenus99}.

On the densely-packed branch, our data indicate that the scaling of $n_l$ and $C_l$ fits well the formula $n_{l}  \sim a+bL^{-2}$ and
$C_{l}  \sim a+bL^{-2}$, while for $n_b$ and $C_b$, there's no detectable finite-size dependence.
The asymptotic behaviour of $B$ for $n \rightarrow 0$ agrees with the prediction $ B (n \rightarrow 0) = 35.70 (2)$~\cite{JacobsenVannimenus99}.

On the fully-packed branch, we have $\mathcal{N}_b=|V|$ and so the expectation of the arithmetic mean of the loop lengths is simply $2/n_l$.
Our data for $n_l$ fit the formula $n_l \sim a+bL^{-2}$, however, for $n=2$ the dominant correction exponent appears to be $-2.20(4)$,
most probably arising from logarithmic corrections. No detectable finite-size dependence is found for $C_l$. For $n=2$,
the estimates for $n_{l0}$, $C_{l0}$ and $B$ confirm the prediction $n_l=1/9$, $C_l=1/9+1/135$, and $B=36$.
We also determined the asymptotic value $B (n \rightarrow 0) = 30.04 (2)$, in good agreement with the exact value $30.0344\ldots$.

\begin{table}[htb]
  \centering
  \caption{\label{Constants table} Numerically determined results $n_{b0}$, $C_{b0}$, $n_{l0}$, $C_{l0}$,
  and $B= n n_{b0}/n_{l0}$.
  }
  \medskip
      {\footnotesize
	\begin{tabular}{|l|l|lllll|}
	  \hline
	  Branch &$n$   & $n_{b0}$        & $C_{b0}$       & $n_{l0}$       & $C_{l0}$         & $B$ \\
	  \hline
      \multirow{4}{*}{Critical}
          & $0.5$       & $0.26646(6) $      & $-2.37(4)$     & $0.018644(5)$     & $0.011(4)$       & $ 7.146(4)$ \\
 	  & $1.0$       & $0.49998(2) $      & $ 0.41(4)$     & $0.033664(4)$     & $-0.0072(3)$     & $14.852(3)$ \\
	  & $1.5$       & $0.72953(1) $      & $ 4.8(1)$      & $0.046243(3)$     & $0.078(2)$       & $23.664(3)$ \\
 	  & $2.0$       & $1.114837(3)$      & $ 1.789(2)$    & $0.057418(2)$     & $0.0553(2)$      & $38.832(2)$ \\
	  \hline
      \multirow{3}{*}{DPL}
 	  & $1.5$       & $1.395506(2)$      & $0.9760(3)$    & $0.0475624(4)$    & $0.03978(2)$     & $44.0108(6)$ \\
 	  & $1.0$       & $1.499999(2)$      & $0.7500(1)$    & $0.0352504(4)$    & $0.02999(2)$     & $42.5527(6)$ \\
          & $0.5$       & $1.579679(2)$      & $0.5960(2)$    & $0.0199245(6)$    & $0.01798(2)$     & $39.642(2) $ \\
      \hline
      \multirow{6}{*}{FPL}
          & $0.1$       & $2$                & $0$            & $0.006527(4)$     & $0.00640(2)$     & $30.642(9) $ \\
 	  & $1.0$       & $2$                & $0$            & $0.057668(2)$     & $0.05242(3)$     & $34.6813(4)$ \\
 	  & $1.25$      & $2$                & $0$            & $0.070680(1)$     & $0.06484(4)$     & $35.3707(2)$ \\
  	  & $1.5$       & $2$                & $0$            & $0.083677(2)$     & $0.07851(5)$     & $35.8521(4)$ \\
   	  & $1.75$      & $2$                & $0$            & $0.096989(2)$     & $0.0952(2) $     & $36.0866(4)$ \\
          & $2.0$       & $2$                & $0$            & $0.111111(2)$     & $0.1185(1) $     & $36.0000(3)$ \\
      \hline
	\end{tabular}
      }
\end{table}

\subsubsection{\texorpdfstring{Scaling of $\<\mathcal{L}_1\>$ and $\<\mathcal{L}_2\>$}{Scaling of the loop-length observables}}
\begin{figure}[ht]
\begin{center}
\includegraphics[angle=-90,scale=0.3]{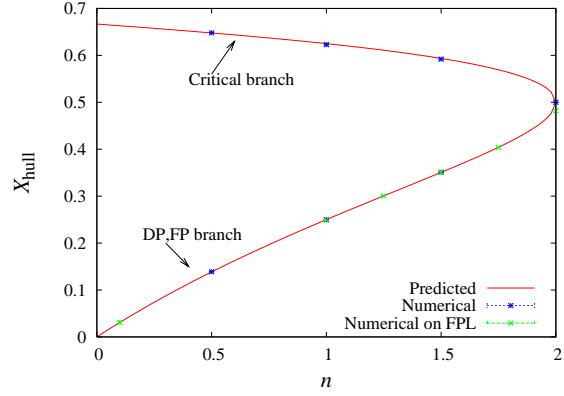}
\caption{
  Numerically determined scaling dimension $X_{\rm loop}$ plotted with the exact expression \eqref{hull exponent} for $X_{\rm hull}$, as a function of $n$.
}
\label{hull plot}
\end{center}
\end{figure}
Standard finite-size scaling arguments predict that
\begin{align}
\<\mathcal{L}_1\> &\sim L^{y_{\rm loop}}\\
\<\mathcal{L}_2\> &\sim L^{2\,y_{\rm loop}-2}
\end{align}
where $y_{\rm loop} = 2 - X_{\rm loop}$ is the fractal dimension characterizing loop length, and $X_{\rm loop}$ is the corresponding scaling dimension.
It was argued in~\cite{SaleurDuplantier87} that for the critical loop model we have $X_{\rm loop}=X_{\rm hull}$, where $X_{\rm hull}$ is the hull scaling dimension
\begin{equation}
  X_{\rm hull} = 1-\frac{2}{g},
  \label{hull exponent}
\end{equation}
and $g$ is related to $x$ and $n$ as in \eqref{n-g relation} and \eqref{x-g relation}.
It was argued in~\cite{KondevDeGierNienhuis96} that $X_{\rm loop}=X_{\rm hull}$ also holds on the fully-packed branch.

On the critical branch, $X_{\rm loop}=X_{\rm hull}$ was verified for $1\le n\le2$ in the Monte Carlo study presented in~\cite{DengGaroniGuoBloeteSokal07}.
Table~\ref{critical branch exponents table} confirms this result, and shows that it extends to $n< 1$.
Since the expression \eqref{hull exponent} for $X_{\rm hull}(g)$ as a function of $g$ should be universal, we expect that if we insert the DPL expression for $g$ into~\eqref{hull exponent} then
we would again have $X_{\rm loop}=X_{\rm hull}$. Table~\ref{DPL branch exponents table} shows that this is indeed the case,
and Table~\ref{FPL branch exponents table} shows that the result also holds on the fully-packed branch.
In Fig.~\ref{hull plot} we plot our numerical estimates of $X_{\rm loop}$ together with the exact result for $X_{\rm hull}$ for all three branches. The agreement is clearly excellent.
The small deviation of the FPL estimate at $n=2$ is presumably due to the presence of logarithmic corrections to scaling.

\subsubsection{\texorpdfstring{Scaling of $\<\mathcal{G}_1\>$ and $\<\mathcal{G}_2\>$}{Scaling of the face-size observables}}
\begin{figure}[ht]
\begin{center}
\includegraphics[angle=-90,scale=0.3]{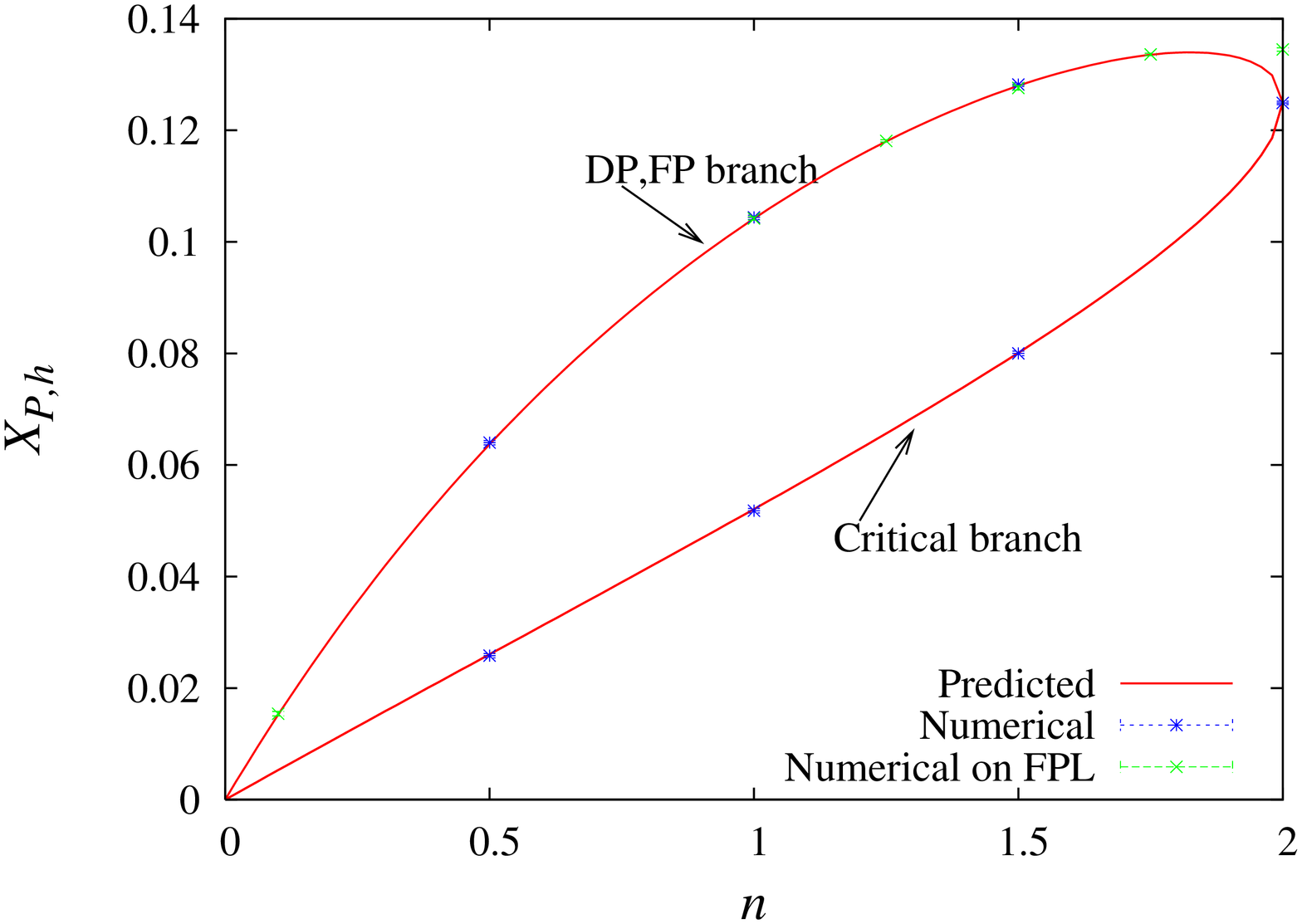}
\caption{
  Numerically determined scaling dimension $X_{\rm face}$ plotted with the exact expression \eqref{magnetic exponent} for $\XhPotts$, as a function of $n$.
}
\label{g2 plot}
\end{center}
\end{figure}
Analogously to the previous case for loop length, we expect that
\begin{align}
\<\mathcal{G}_1\> &\sim L^{y_{\rm face}}\\
\<\mathcal{G}_2\> &\sim L^{2\,y_{\rm face}-2}
\end{align}
where $y_{\rm face} = 2 - X_{\rm face}$ is the fractal dimension characterizing face size, and $X_{\rm face}$ is the corresponding scaling dimension.
It was argued in~\cite{DuplantierSaleur89} that for both the critical and densely-packed loop models we have $X_{\rm face}=\XhPotts$,
where $\XhPotts$ is the magnetic scaling dimension of the $q$-state Potts model with $q=n^2$,
whose expression in terms of the Coulomb gas coupling $g$ is known~\cite{Nienhuis84} to be
\begin{equation}
\XhPotts  = \frac{(6-g)(g-2)}{8g},
\label{magnetic exponent}
\end{equation}
and $g$ is related to $x$ and $n$ as in \eqref{n-g relation} and \eqref{x-g relation}.

On the critical branch, $X_{\rm face}=\XhPotts$ was verified for $1\le n\le2$ in the Monte Carlo study presented in~\cite{DengGaroniGuoBloeteSokal07}.
Table~\ref{critical branch exponents table} confirms this result, and shows that it extends to $n < 1$,
and Table~\ref{DPL branch exponents table} verifies it in the densely-packed phase.
It is not clear {\em a priori} that this relationship should also hold in the fully-packed branch, but Table~\ref{FPL branch exponents table} shows that it does.
In Fig.~\ref{g2 plot} we plot our numerical estimates of $X_{\rm face}$ together with the exact result for $\XhPotts$ for all three branches. The agreement is clearly excellent.
The deviation of the FPL estimate at $n=2$ is presumably due to the presence of logarithmic corrections-to-scaling.

\subsubsection{\texorpdfstring{Scaling of $\chi_{\rm Ising}$}{Scaling of the dual Ising susceptibility}}
\begin{figure}[ht]
\begin{center}
\includegraphics[angle=-90,scale=0.3]{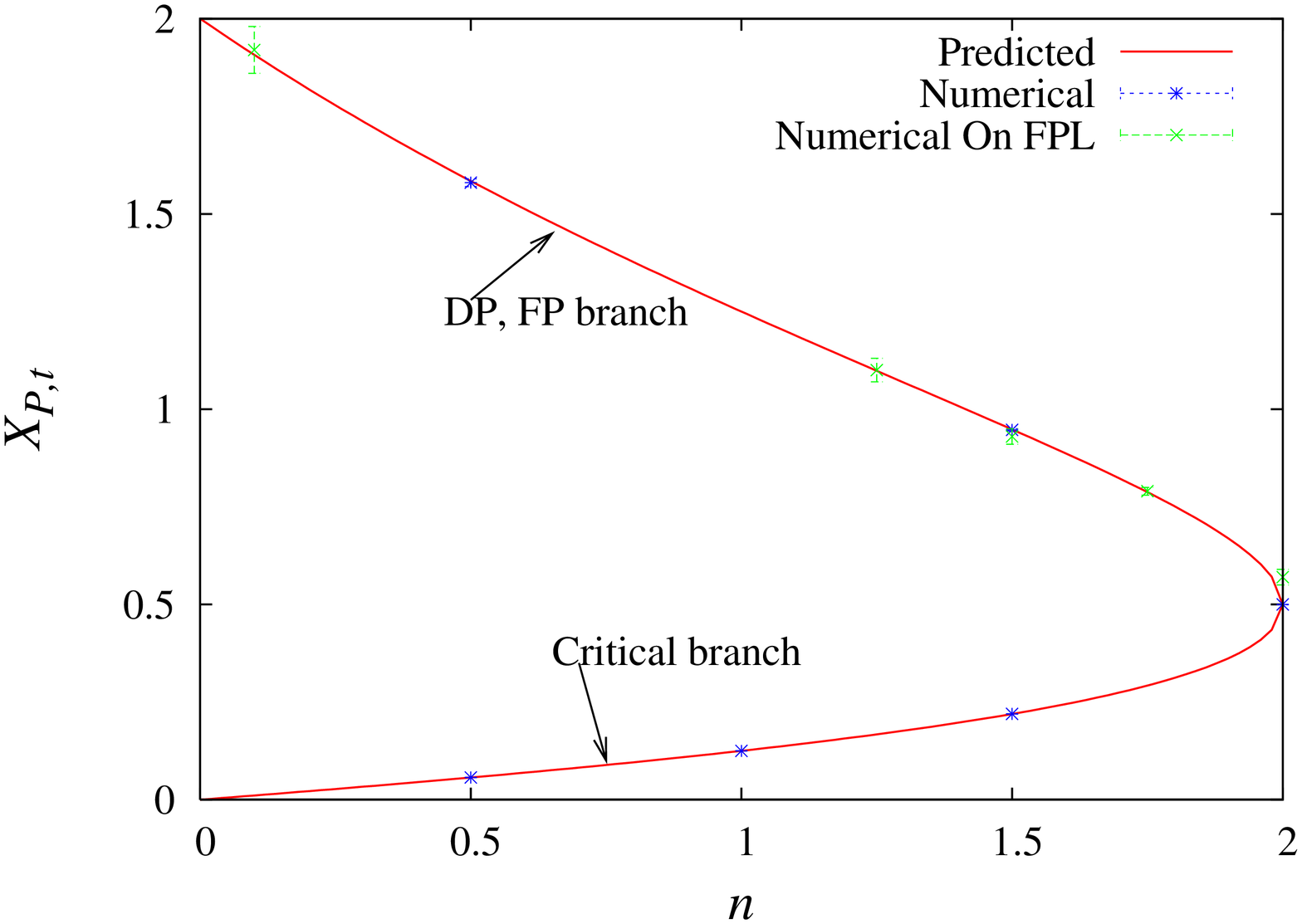}
\caption{
  Numerically determined scaling dimension $X_{\rm Ising}$ plotted with the exact expression \eqref{thermal exponent} for $\XtPotts$, as a function of $n$.
}
\label{chi plot}
\end{center}
\end{figure}
It is natural to expect that
\begin{equation}
  \chi_{\rm Ising} \sim L^{2-2 X_{\rm Ising}}
\end{equation}
for some scaling dimension $X_{\rm Ising}$.
It is not {\em a priori} obvious what the form of $X_{\rm Ising}$ should be, however it was observed in~\cite{DengGaroniGuoBloeteSokal07} for $1\le n\le 2$ that on the critical branch we have
$X_{\rm Ising}=\XtPotts$, where $\XtPotts$ is the thermal scaling dimension of the $q$-state Potts model with $q=n^2$, whose expression in terms of the Coulomb gas coupling $g$ is known~\cite{Nienhuis84} to be
\begin{equation}
\XtPotts  = \frac{6}{g}-1,
\label{thermal exponent}
\end{equation}
and where $g\in[4,6]$ is related to $x$ and $n$ as in \eqref{n-g relation}.
Table~\ref{critical branch exponents table} confirms this result, and shows that it extends to $n < 1$.
One would expect that the relationship would continue into the densely-packed phase, with $g\in[2,4]$, and Table~\ref{DPL branch exponents table} shows that this is indeed the case.
It is not entirely obvious what the behaviour should be on the fully-packed branch, but Table~\ref{FPL branch exponents table} shows that it simply coincides with that of the densely-packed branch.
We note that on the densely-packed and fully-packed branches, when $x<\sqrt{2}$ we have $2-2X_{\rm Ising}<0$ and so in these cases we estimated $X_{\rm Ising}$ from an ansatz of the form
\begin{equation}
  \chi_{\rm Ising} = a + b\,L^{2-2 X_{\rm Ising}} + \ldots
\end{equation}
in which the $L^{2-2 X_{\rm Ising}}$ term is sub-dominant.

In Fig.~\ref{chi plot} we plot our numerical estimates of $X_{\rm Ising}$ together with the exact result for $\XtPotts$ for all three branches. The agreement is clearly excellent.
We note that there is no data point for $n=1$ on the densely-packed branch, since $(x,n)=(1,1)$ simply corresponds to site percolation on the dual triangular lattice
(the $+ (-)$ spins are regarded as occupied (vacant) sites).
In this case, the nontrivial dependence of $\chi_{\rm Ising}$ on the system size vanishes; i.e. the amplitude $a_0$ in~(\ref{FSS ansatz}) is identically zero.

\subsubsection{\texorpdfstring{Scaling of $\chi_{\rm stag}$}{Scaling of the staggered dual Ising susceptibility}}
\begin{figure}[ht]
  \begin{center}
    \includegraphics[angle=-90,scale=0.3]{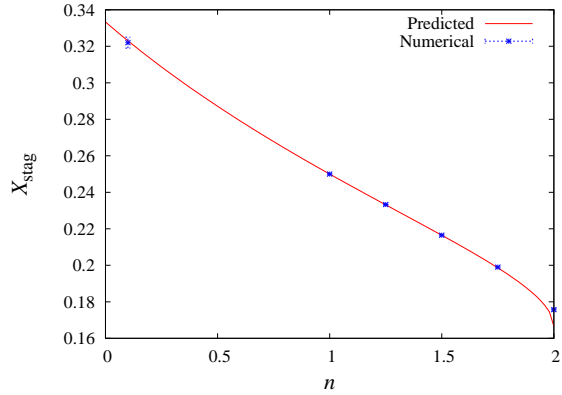}
    \caption{Numerically determined scaling dimension $X_{\rm stag}$ as a function of $n$ in the fully-packed phase, plotted with the conjectured exact expression $X_{\rm stag}=2/3g$.
    }
    \label{staggered plot}
  \end{center}
\end{figure}
On the critical branch, one might intuitively expect that $\chi_{\rm stag}$ converges to a constant as $L$ increases,
since the symmetry of the sublattices should cause the staggered magnetization to cancel out for ferromagnetic models.
Indeed, our numerical data show that on the critical branch $\chi_{\rm stag}$ is well described by the simple FSS ansatz
\begin{equation}
  \chi_{\rm stag} = a + b\,L^{-X_{\rm stag}}.
\end{equation}
Perhaps surprisingly, our data also strongly suggest that we can in fact make the identification $X_{\rm stag}=\XtSubPotts$
where $\XtSubPotts$ is the second thermal scaling dimension of the $q$-state Potts model with $q=n^2$, whose expression in terms of the Coulomb gas coupling $g$ is~(\ref{subpotts_exponent}). See Table~\ref{critical branch exponents table}.

In the densely-packed phase, when $n=1$ it is clear that $\chi_{\rm stag}$ is simply a constant which displays no finite-size dependence,
since the model corresponds to site percolation on the triangular lattice in this case.
It is not obvious that this behaviour should persist when $n\neq1$, however our simulations strongly suggest that this is indeed the case:
we find empirically that there is no finite-size dependence of $\chi_{\rm stag}$ for any $0<n\le2$ on the densely-packed branch.

In the fully-packed phase, the $n=1$ case corresponds to the zero-temperature antiferromagnetic Ising model on the dual triangular lattice, which is known to be critical,
and $\chi_{\rm stag}$ is its order parameter.
It follows that for $n=1$ we have
\begin{equation}
\chi_{\rm stag}\sim L^{2 - 2 X_{\rm stag}}
\label{FPL staggered ansatz}
\end{equation}
with $X_{\rm stag}=1/4$. Our numerical results show convincingly that in fact \eqref{FPL staggered ansatz} holds for all $n\le2$.
However, we were not able to identify the exponent $X_{\rm stag}$ in terms of other known exponents.
Instead, we fitted our empirical estimates of $X_{\rm stag}$ to a simple Coulomb-gas ansatz
\begin{equation}
X(g) = \frac{a g^2 +b g +c}{d g}
\label{fit_exponent}
\end{equation}
with $a, b, c, d$ unknown integers. We found that $X_{\rm stag}$ is well described by the simple formula (see Fig.~\ref{staggered plot})
\begin{equation}
  X_{\rm stag}= \frac{2}{3\,g}.
  \label{X stag conjecture}
\end{equation}
Based on the agreement presented in Table~\ref{DPL branch exponents table}, we conjecture that this formula is in fact exact.

\subsubsection{\texorpdfstring{Scaling of $\<\mathcal{T}\>$}{Scaling of the return time}}
\begin{figure}[ht]
  \begin{center}
    \includegraphics[angle=-90,scale=0.3]{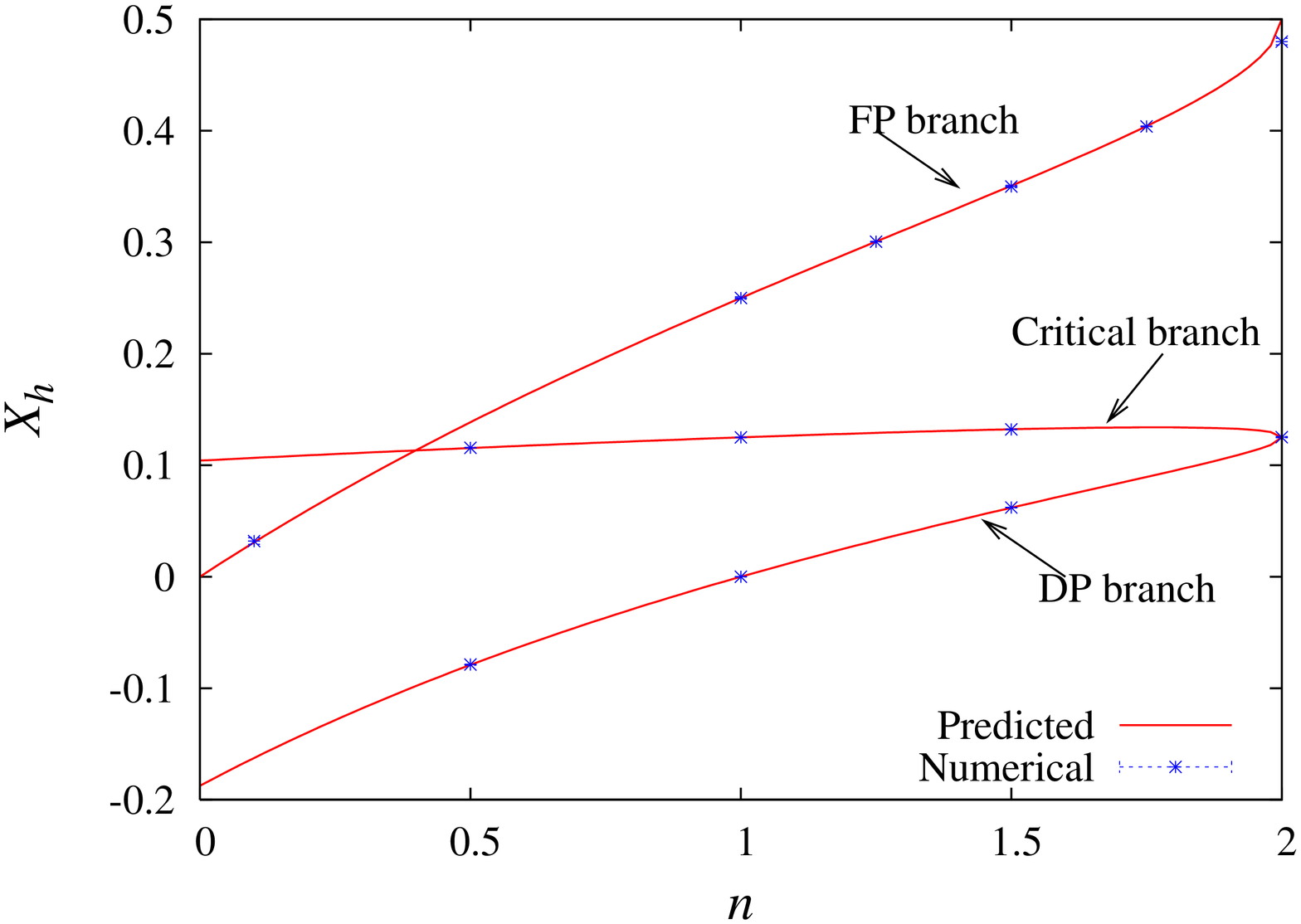}
    \caption{
      Numerically determined scaling dimension $X_{\rm worm}$ plotted with the exact prediction for $\XhOn$ as a function of $n$.
    }
    \label{return time plot}
  \end{center}
\end{figure}
It is natural to expect that
\begin{equation}
\<\mathcal{T}\> \sim L^{2\,y_{\rm worm}-2}
\end{equation}
where $y_{\rm worm} = 2 - X_{\rm worm}$ is a fractal dimension characterizing the time of return to the Eulerian subspace $\mathcal{C}(G)\times V\subseteq\mathcal{S}(G)$ in
the connectivity-checking version of the worm dynamics.

Although $\mathcal{T}$ is inherently an observable on the state space of the worm dynamics, $\mathcal{S}(G)$, rather than on the state space of the loop model, $\mathcal{C}(G)$,
its scaling is linked to the loop model in a fundamental way. In fact, we find numerically that $X_{\rm worm}=\XhOn$ for all three branches and for all $n\le2$,
where $\XhOn$ is the magnetic scaling dimension of the $O(n)$ loop model.
We note that, as shown in~\cite{ProkofevSvistunov01,DengGaroniSokal07c}, on the critical branch of
the $n=1$ model $\<\mathcal{T}\>$ is equal to the susceptibility of the Ising model whose high-temperature graphs are sampled by the worm dynamics.
Therefore, in this special case we have a sound theoretical argument that $X_{\rm worm}=\XhOn$, so the general result is not completely unexpected.

On the critical and densely-packed branches, the exact expression~\cite{Nienhuis82,BatchelorBloete89} for $\XhOn$ is given by
\begin{equation}
  \XhOn  = 1 - \frac{2}{g} - \frac{3}{32}g,
\end{equation}
where $g$ is related to $x$ and $n$ as in \eqref{n-g relation} and \eqref{x-g relation}, while on the fully-packed branch it is~\cite{BloeteNienhuis94,BatchelorSuzukiYung94,KondevDeGierNienhuis96}
\begin{equation}
  \XhOn  = 1 - \frac{2}{g},
\end{equation}
with $g\in[2,4]$ related to $n$ as in \eqref{n-g relation}.
Tables~\ref{critical branch exponents table}, \ref{DPL branch exponents table} and~\ref{FPL branch exponents table} show a comparison between $X_{\rm worm}$ and $\XhOn$, and
in Fig.~\ref{return time plot} we plot our numerical estimates of $X_{\rm worm}$ together with the exact result for $\XhOn$ for all three branches. The agreement is clearly excellent.

Based on these results for the honeycomb lattice, it appears that the identity $X_{\rm worm}=\XhOn$ is a generic property of the worm algorithms we present.
Preliminary results for the hydrogen-peroxide lattice further support this conjecture; these results will be reported in detail elsewhere.
For this reason, it seems in a certain sense that the worm algorithm provides the {\em natural} dynamics for the $O(n)$ loop model.
The relation $X_{\rm worm}=\XhOn$ also has some important practical consequences, since it provides a simple way of directly measuring $\XhOn$.
As one application, it can be used to obtain an accurate estimate of the curve $X_h(n)$ as a continuous function of $n$ in three dimensions, where no exact results are currently known.

The identification $X_{\rm worm}=\XhOn$ also has implications for the efficiency of the worm dynamics.
For $n<1$, the exponent $2-2\XhOn$ governing $\langle\mathcal{T}\rangle$ is greater than 2 in the densely-packed phase,
implying that it takes a time larger than order volume between visits to the Eulerian subspace.
The efficiency of the worm dynamics will therefore suffer in this region. By contrast, on the critical branch one has $2-2\XhOn\in(1.73,1.8)$ for all $0\le n\le 2$, while on the fully-packed
branch $2-2\XhOn$ decreases monotonically with $n\in[0,2]$ from $2$ to $1$.

\begin{remark}
  Although we have not proved that Algorithm~\ref{fpl algorithm} remains ergodic at $n=0$, we used it to perform simulations of the uniform Hamiltonian cycle model, and we found that $X_{\rm worm}=0=\XhOn(n=0)$.
  This provides some modest evidence that the conjecture $X_{\rm worm}=\XhOn$ in fact extends to $n=0$ when $x=\infty$, and also that Algorithm~\ref{fpl algorithm} may indeed be ergodic.
\end{remark}

\begin{remark}
  For both the critical branch and densely-packed branch, one has the choice of using either the simple worm dynamics, given by \eqref{worm P}, or the rejection-free worm dynamics \eqref{rejection-free worm P}.
  Perhaps unsurprisingly, both versions have the same value of $X_{\rm worm}$.
  On the fully-packed branch, only the rejection-free dynamics can be used, since the simple dynamics is non-ergodic.
\end{remark}

\begin{remark}
  When $n > 1$, one can also consider $X_{\rm worm}$ for the colouring version of the worm dynamics.
  On the critical branch, we simulated both versions and measured $\mathcal{T}$ in both cases.
  There appears to be no reason, {\em a priori}, for $\langle\mathcal{T}\rangle$ in the two cases to be related in any simple way, or even to share the same value of $X_{\rm worm}$.
  However, empirically, we found that on the critical branch the exponent $X_{\rm worm}$ governing $\langle\mathcal{T}\rangle$ agrees, within error bars, for the two versions.
  On the densely-packed branch, by contrast, we found $X_{\rm worm}=0.053(2)$ for the colouring version when $n=1.5$, which appears to disagree with the value $X_{\rm worm}=0.0620(3)$ for the
  connectivity-checking version reported in Table~\ref{DPL branch exponents table}.
  Note that this implies the colouring version has the larger mean return time $\langle \mathcal{T} \rangle$, since $2-2X_{\rm worm}$ is larger in this case.
  Since we only simulated one value of $n>1$ on the DP branch, this is the only point where a comparison between the two versions can be made.
  We did not perform any systematic numerical tests of the colouring algorithm in the fully-packed phase,
  since our preliminary results suggested it was significantly less efficient than the connectivity-checking version.
\end{remark}

\subsubsection{\texorpdfstring{Scaling of $\mathbb{P}(\mathcal{T}=t)$}{Scaling of the distribution of the return time}}
In order to better understand the behaviour of $\mathcal{T}$, we computed its histogram and thereby estimated its distribution $\mathbb{P}(\mathcal{T} = t)$.
One expects that $\mathbb{P}(\mathcal{T}=t)$ obeys a scaling law for large $t$ in a critical system. More precisely, we expect that at finite volume we have
\begin{equation}
\mathbb{P}(\mathcal{T}=t) \sim t^{-\tau}f\left(t/L^y\right),
\label{distribution ansatz}
\end{equation}
for some choice of exponents $\tau$ and $y$, and some scaling function $f$.
Note that the ansatz \eqref{distribution ansatz} implies that $\<\mathcal{T}^k\>\sim L^{(k+1-\tau)y}$ to leading order as $L\to\infty$.

The distribution of the loop length was studied in~\cite{KondevDeGierNienhuis96} and found to be of the form \eqref{distribution ansatz}.
We have computed the distributions of $\mathcal{L}_2$ and $\mathcal{G}_2$ in the present work, and also found excellent fits to \eqref{distribution ansatz}.
In addition, distributions of the form \eqref{distribution ansatz} are known to govern the cluster-size of the random cluster model~\cite{DengZhangGaroniSokalSportiello10}.
For all these cases, the exponent $y$ appearing in \eqref{distribution ansatz} coincides with the fractal dimension of the corresponding geometric objects;
in particular we have $y=y_{\rm face} = 2-\XhPotts$ and $y=y_{\rm loop} = 2-X_{\rm hull}$ for the face-size and loop-size cases, respectively.
In such cases the two parameters $y$ and $\tau$ appearing in \eqref{distribution ansatz} can be related by the following argument.
Consider a generic observable $\mathcal{A}$ characterizing a geometric property (cluster-size, loop-size, face-size\ldots)
with fractal dimension $y_{\mathcal{A}}$. Since a given object has scale $L^{y_{\mathcal{A}}}$, the probability of picking one at random should be $\sim L^{\rm y_{\mathcal{A}}}/L^d$ so that we expect
$\<\mathcal{A}^k\>\sim L^{y_{\mathcal{A}}-2}\,L^{k\,y_{\mathcal{A}}}=L^{(k+1)y_{\mathcal{A}}-2}$. Combining this latter result with the fact that $\<\mathcal{A}^k\>\sim L^{(k+1-\tau)y}$, and
assuming $y=y_{\mathcal{A}}$, it follows that $\tau=2/y$.

\begin{figure}[htb]
  \begin{center}
    \includegraphics[angle=-90,scale=0.3]{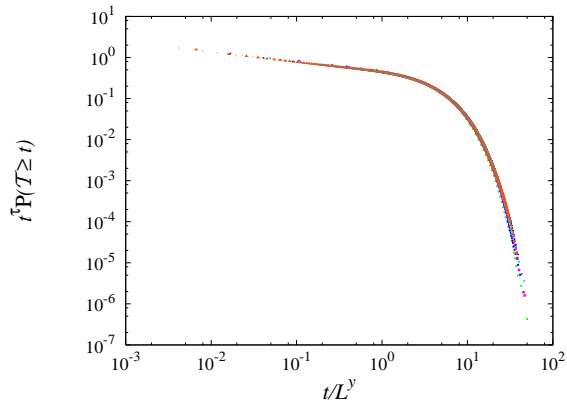}
    \caption{Finite-size-scaling plot showing $t^{\tau}\,\mathbb{P}(\mathcal{T}\ge t)$ versus $t/L^y$ for $n=1$ on the critical branch, with system sizes $L=12,24,48,96,192,384,768$.}
    \label{scaled return time distribution plot}
  \end{center}
\end{figure}

Table~\ref{distribution exponents} lists our fits for $y$ and $\tau$, together with $y_{\rm worm}$ for ease of comparison, and
Fig.~\ref{scaled return time distribution plot} shows a finite-size scaling plot of $\mathbb{P}(\mathcal{T}\ge t)$ for $n=1$ on the critical branch.
\begin{table}
  \centering
  \caption{\label{distribution exponents}Estimated values of the scaling exponents, $y$ and $\tau$, appearing in the ansatz \eqref{distribution ansatz} for the distribution $\mathbb{P}(\mathcal{T} = t)$.
    Values for the critical, densely-packed, and fully-packed branches are shown.
  }
  \medskip
      {\footnotesize
	\begin{tabular}{|l|lll|lll|lll|}
	  \hline
	  \multicolumn{1}{|c}{} & \multicolumn{3}{c|}{Critical} & \multicolumn{3}{c|}{Densely-packed} & \multicolumn{3}{c|}{Fully-packed} \\
	  \hline
	  $n$    & $\tau$     & $y$         & $y_{\rm worm}$ & $\tau$     & $y$        & $y_{\rm worm}$ & $\tau$     & $y$         & $y_{\rm worm}$\\
	  \hline
	  $0.1$  & \---       & \---        & \---           & \---       & \---       & \---           & $1.166(6)$ & $2.32(1)$   & $1.968(4)$    \\
 	  $0.5$  & $1.286(4)$ & $2.479(4)$  & $1.8843(3)$    & $1.044(5)$ & $2.257(7)$ & $2.0788(4)$    & \---       & \---        & \---          \\
 	  $1$    & $1.248(3)$ & $2.325(3)$  & $1.8750(2)$    & $1.077(4)$ & $2.164(6)$ & $2.0000(1)$    & $1.364(3)$ & $2.360(5)$  & $1.7502(4)$   \\
 	  $1.5$  & $1.223(4)$ & $2.234(4)$  & $1.8678(2)$    & $1.108(5)$ & $2.103(8)$ & $1.9380(3)$    & $1.471(4)$ & $2.458(7)$  & $1.6493(3)$   \\
 	  $2$    & $1.174(4)$ & $2.120(4)$  & $1.8749(5)$    & \---       & \---       & \---           & $1.602(5)$ & $2.62(1)$  & $1.520(3)$    \\
	  \hline
	\end{tabular}
      }
\end{table}
Contrary to the observables discussed above, our numerical results show clearly that for the observable $\mathcal{T}$,
the exponent $y$ appearing in \eqref{distribution ansatz} is {\em not} equal to the corresponding fractal dimension $y_{\rm worm}$.
Although the combined exponent $(2-\tau) y = 2-2\XhOn$ governing the mean $\langle\mathcal{T}\rangle$ is universal, it remains an open question whether or not $\tau$ and $y$ are themselves universal.
We tried to fit the values of $y$ in Table~\ref{distribution exponents} to the ansatz~\eqref{fit_exponent} with $d=8, 12$, and 16,
but we did not obtain a meaningful expression within the estimated statistical errors.

\subsection{\texorpdfstring{$O(n)$ loop model for $n>2$ }{n > 2 }}
Until quite recently, it was generally believed~\cite{KunzWu88} that the (two-dimensional) $O(n)$ loop model, \eqref{loop measure}, does not exhibit a phase transition when $n>2$.
Using numerical transfer matrix methods, however, strong evidence was found in~\cite{GuoBloeteWu00} that a line of critical points does in fact exist for all $n>2$.
Furthermore, the results presented in~\cite{GuoBloeteWu00} also suggest that this phase transition falls into the three-state Potts universality class.
This can be understood by noting that the $n\to\infty$ loop model is equivalent to the hard-hexagon model~\cite{DomanyMukamelNienhuisSchwimmer81}.

\begin{figure}[ht]
  \begin{center}
    \includegraphics[angle=-90,scale=0.3]{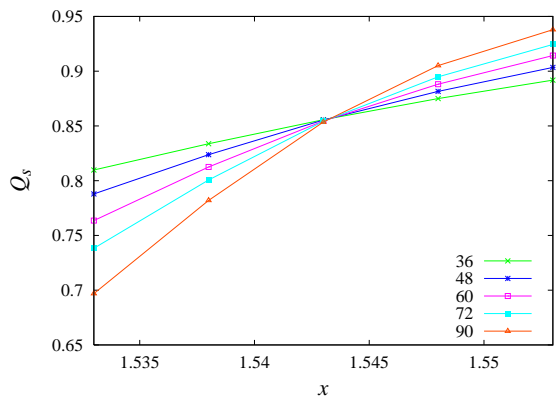}
    \caption{The Binder cumulant $Q_s$ for $n=10$.
    }
    \label{Binder_q}
  \end{center}
\end{figure}

We performed Monte Carlo simulations of the $n=3$ and $n=10$ loop models, using the worm algorithm presented in Section~\ref{Worm algorithms}.
We computed the staggered magnetization on the dual triangular lattice, which allowed us to then compute the staggered susceptibility and the dimensionless ratio $Q_s$.
By studying $Q_s$ we could accurately locate the critical point. See Fig.~\ref{Binder_q} and Table~\ref{n>2 table}. By studying both $\chi_{\rm stag}$ and $Q_s$ we also estimated
the exponents $X_t$ and $X_h$. See Table~\ref{n>2 table}. Our numerical values for $X_t$ and $X_h$ are entirely consistent with a transition in the 3-state Potts universality class,
which has $X_t=4/5=0.8000$ and $X_h=2/15=0.1333$.

For comparison, we note that in~\cite{GuoBloeteWu00} it was estimated that $x_c=1.52(1)$, $X_t=0.80(1)$ and $X_h=0.14(1)$ for $n=10$.
No data for $n=3$ was reported in~\cite{GuoBloeteWu00}, however their quoted values for $n=4$ were $X_t=0.76(5)$ and $X_h=0.1(1)$.
A number of different $n$ values were reported in~\cite{GuoBloeteWu00}, and the accuracy of their estimated exponents was found to increase with $n$.
By contrast, we found the computational effort required for the $n=3$ and $n=10$ cases to be comparable.
\begin{table}
  \centering
  \caption{\label{n>2 table} Estimates of $X_t$, $X_h$ and $x_c$ for the $O(n)$ loop model with $n=3,10$.
    For comparison, note that for the critical 3-state Potts we have $X_t=4/5$ and $X_h=2/15$.
  }
  \medskip
      {\footnotesize
	\begin{tabular}{|l|lll|}
	  \hline
	  $n$    & $x_c$        & $X_t$     & $X_h$     \\
	  \hline
	  $3$    & $6.822(7)$   & $0.81(3)$ & $0.132(3)$ \\
	  $10$   & $1.5430(2)$  & $0.83(4)$ & $0.135(4)$ \\
	  \hline
	\end{tabular}
      }
\end{table}

\section{Discussion}
\label{discussion}
We have presented a Markov-chain Monte Carlo algorithm of {\em worm} type that correctly simulates the $O(n)$ loop model on any bipartite cubic graph, for any
$n\in(0,\infty)$ and $x\in(0,\infty]$, and we have proved rigorously that the algorithm is ergodic and has the correct stationary distribution.
We have then applied this algorithm to the honeycomb-lattice loop model.
Comparing our numerical results when $n\le2$ with Coulomb gas theory allowed us to identify the exact exponents of a number of fundamental geometric observables,
as well as observables related to dual Ising spins.
Furthermore, we have provided compelling numerical evidence that $X_{\rm worm}=X_h$ in all three branches for all $n\le2$.
This suggests that $X_{\rm worm}$ can be used as an efficient means to estimate $X_h$ in models where no theoretical predictions exist, such as in three dimensions.
For the case $n>2$, we confirmed the existence of a phase transition in the 3-state Potts universality class, which has previously been observed using transfer matrices.
Equipped with our worm algorithm, it is now natural to repeat these studies on bipartite cubic graphs other than the honeycomb-lattice, including to the Hydrogen-peroxide lattice, which
provides a natural three-dimensional generalization, and the $(4\cdot8^2)$ Archimedean lattice, which is the dual of the Union Jack lattice.
These results will be reported elsewhere.

Another natural application of the worm algorithms presented in this article is to the study of antiferromagnetic Potts models on triangulations of the torus.
Since antiferromagnetic models do not display universality, it is of significant interest to study such models on a variety of different lattices.
As discussed in the introduction, it is well known that the honeycomb-lattice FPL model with $n=1$ is equivalent to the zero-temperature triangular-lattice antiferromagnetic Ising model.
While cluster algorithms~\cite{ZhangYang94,CoddingtonHan94} for this model are thought to be non-ergodic,
the worm algorithm presented in Section~\ref{Worm algorithms} provides a provably valid Monte Carlo method; see~\cite{ZhangGaroniDeng09}.

This observation can be generalized in two ways. Firstly, the dual of any bipartite cubic map on the torus is an Eulerian triangulation, and
the worm algorithm from Section~\ref{Worm algorithms} can immediately be applied to study Ising models on these triangulations.
Secondly, as already mentioned in the Introduction, the honeycomb-lattice FPL model with $n=2$ is equivalent to the zero-temperature triangular-lattice 4-state Potts antiferromagnet~\cite{Baxter70},
as well as the zero-temperature kagome-lattice 3-state Potts antiferromagnet.
As also already mentioned, the Wang-Swendsen-Koteck\'y~\cite{WangSwendsenKotecky90} (WSK) cluster algorithm,
which is undoubtedly the current state-of-the-art for simulating antiferromagnetic Potts models, has recently been proved~\cite{MoharSalas09,MoharSalas10} to be non-ergodic for both of these models.
By contrast, the worm algorithms described in Section~\ref{Worm algorithms} have been proved to be ergodic, and they can be easily applied to the corresponding loop models,
in order to then study these Potts antiferromagnets.
To do so, it is sufficient to simply augment the usual worm steps with an additional step that randomly assigns an alternating colouring (red, blue, red,\ldots) to the edges of each cycle.
Each cycle can be coloured in precisely two ways. By interpreting the remaining edges as green,
this defines a new transition matrix on 3-edge colourings of the honeycomb lattice, given by $2^{-c(A)}P_{G,n=2}[(A,u,v)\to(A\symdif uu',u',v)]$.
Since $\pworminfty$ is in detailed balance with $\phi_{G,n}(A)\propto n^{c(A)}$, at $n=2$ this new transition matrix uniformly samples 3-edge colourings.
But since the kagome lattice is the medial graph of the honeycomb lattice, it immediately follows that this new transition matrix in fact uniformly samples 3-vertex colourings of the kagome lattice
(i.e. simulates the zero-temperature kagome-lattice Potts antiferromagnet). The $4$-state model can be treated in a similar way.
Finally, we note that these mappings from $n=2$ loop models to zero-temperature $q=3$ and $q=4$ state Potts models hold quite generally.
In particular, the 4-state antiferromagnetic Potts model on a variety of two-dimensional triangulations can be studied using such worm algorithms.
All of these possibilities we leave for future work.

\section*{Acknowledgments}
This work was supported in part by the National Nature Science Foundation of China under Grant No. 10975127, the Anhui Provincial Natural Science Foundation under Grant No. 090416224, 
and the Chinese Academy of Sciences. 
The authors wish to thank an anonymous referee for some very valuable comments.
YD and TMG are indebted to Jan de Gier, Alan Sokal and Jes\'us Salas for helpful discussions.

\bibliographystyle{elsarticle-num}
\bibliography{garoni}

\end{document}